\documentclass[aps,pra,preprint,superscriptaddress, twocolumn, 10 pt]{revtex4-1}
\usepackage[utf8]{inputenc}
\usepackage{graphicx}
\usepackage{amssymb}
\usepackage{amsmath}
\usepackage{amstext}
\usepackage{amsthm}
\usepackage{amsfonts}
\usepackage{color}
\usepackage{xcolor}

\newlength{\myleftlen}

\def\beq{\begin{equation}}
\def\eeq{\end{equation}}
\def\beqa{\begin{eqnarray}}
\def\eeqa{\end{eqnarray}}

 \newcommand{\expval}[1]{\langle#1\rangle} %<"cosa">

\begin{document}

\title{Semiclassical approach to finite temperature quantum annealing with trapped ions}
\author{David Ravent\'os}
\email{David.Raventos@icfo.eu}
\affiliation{ICFO-Institut de Ciencies Fotoniques, The Barcelona Institute of Science and Technology,
 08860 Castelldefels (Barcelona), Spain}
\author{Tobias Gra{\ss}} 
\affiliation{Joint Quantum Institute, University of Maryland, College Park, MD 20742, U.S.A.}
\author{Bruno Juli\'a-D\'iaz} 
\affiliation{Departament de Física Qu\`antica i Astrof\'isica, Facultat 
de Física, 
Universitat de Barcelona, Barcelona 08028, Spain}
\affiliation{Institut de Ci\`encies del Cosmos, Universitat de 
Barcelona, ICCUB, 
Martí i Franqu\`es 1, Barcelona 08028, Spain}
\affiliation{ICFO-Institut de Ciencies Fotoniques, The Barcelona Institute of Science and Technology,
 08860 Castelldefels (Barcelona), Spain}
\author{Maciej Lewenstein}
\affiliation{ICFO-Institut de Ciencies Fotoniques, The Barcelona Institute of Science and Technology,
 08860 Castelldefels (Barcelona), Spain}
\affiliation{ICREA, Passeig de Llu\'is Companys, 23, 08010 Barcelona, Spain}

\begin{abstract}
Recently it has been demonstrated that an ensemble of trapped ions may 
serve as a quantum annealer for the number-partitioning problem [Nature 
Comm. DOI: 10.1038/ncomms11524]. This hard computational problem may be 
addressed employing a tunable spin glass architecture. Following the proposal 
of the trapped ions annealer, we study here its robustness against thermal 
effects, that is, we investigate the role played by thermal phonons. 
For the efficient description of the system, we use a semiclassical approach, 
and benchmark it against the exact quantum evolution. The aim is to understand 
better and characterize how the quantum device approaches a solution of, 
an otherwise, difficult to solve NP-hard problem. 
\end{abstract}

\maketitle

\section{Introduction}
\label{sec:intro}

Quantum computers and quantum simulators are nowadays 
becoming a reality thanks to the advances in ion trapping and integrated 
superconducting technology~\cite{2017arXiv171203773A,supercon,Linke3305,Monroe1164,Devoret1169}. 
A possible device which is quickly being developed are 
quantum annealers. Annealing, as opposed to quenching, is a method to 
produce the ground state of a target Hamiltonian by slowly deforming/adjusting 
a well-known ground state of a different Hamiltonian. Annealing is 
in fact a concept originating from classical metallurgy, extended in the 1980s 
to classical optimization problems, and known as simulated annealing~\cite{Kirkpatrick671,cerny1985}. 
In the current quantum versions, quantum annealing is very much analogous to 
quantum adiabatic computing, but is typically targeted towards the classical 
optimization problems. The idea is to add a simple, noninteracting, but noncommuting 
term to the original classical Hamiltonian. This simple additional term should 
dominate the system at the initial time, so that the ground state will be easy to 
find, since it will correspond to a non-interacting system. The non-commuting 
nature of the additional term ensures that the initial and target ground states
are not symmetry protected. Then, the additional term is adiabatically removed 
and the ground state is expected to go slowly from the initial one to the 
one of the Hamiltonian of interest~\cite{Brooke779,PhysRevE.58.5355,Farhi472}, see also 
the recent review~\cite{RevModPhys.90.015002}. 
This scheme is nowadays plausible with a 
large number of possible platforms, including trapped ions, cavity QED, circuit 
QED, superconducting junctions~\cite{nature10012} and atoms in nanostructures. 
The first commercially 
accessible quantum annealers are in the market~\cite{Rnnow420,Boixo2014,youtube}. 

Since the original proposals~\cite{PhysRevLett.87.257904,PhysRevLett.92.207901} 
trapped ions quantum simulators are the subject of intensive theoretical and 
experimental research. Starting from realization of the simple instances of 
quantum magnetism~\cite{friedenauer08}, they have reached quite a maturity 
in the recent experimental developments  (cf.~\cite{2015arXiv151203559M,
PhysRevA.94.023401,2016arXiv160906429L,PhysRevLett.119.080501,PhysRevLett.118.053001,timecrystal,Zhang2017}). The recent paper by Bollinger's 
group~\cite{2017arXiv171107392S}, in addition to the excellent experimental work, 
contains also an outstanding analysis of quantum dynamics of the relevant Dicke 
model, in which the ions interact essentially with one phononic mode. 

Quantum dynamics in general, and in particular for the Dicke-like ion-phonon models, are very challenging for numerical simulations. Exact treatments are possible for small systems only, so that various approximate methods have to be used. One of them is the truncated Wigner approximation, in which both ionic and phononic operators are replaced by complex numbers, the dynamics becomes ``classical'', and only the initial data mimic the "quantumness" of the problem \cite{TWA}. This approach was used in Ref.~\cite{PhysRevA.96.033607}
to study the quantum non-equilibrium dynamics of spin-boson models. More sophisticated ``mean-field'' approaches decorrelate ions from phonons, but treat at least either ions or phonons fully quantum mechanically -- this approach is in particular analyzed in the present paper. Quantum aspects
of the models in question were studied in the series of papers~\cite{PhysRevA.95.013602,Garttner2017,PhysRevA.94.053637}.

We have considered recently the exact quantum dynamics of few ion systems
to demonstrate the robustness of chiral spin currents in a trapped-ion
quantum simulator using Floquet engineering~\cite{PhysRevA.97.010302}. Our
earlier works include studies of dual trapped-ion quantum simulators as
an alternative route towards exotic quantum magnets~\cite{1367-2630-18-3-033011}, and
studies of ion chains with long range interactions forming ``magnetic loops''~\cite{PhysRevA.91.063612}.
Topological edge states in periodically-driven trapped-ion chains~\cite{PhysRevLett.119.210401}, trapped
ion quantum simulators of Rabi lattice models with discrete gauge symmetry~\cite{2015PhRvA..92a3624N}, and hidden frustrated interactions
and quantum annealing in trapped ion spin-phonon chains~\cite{PhysRevA.93.013625} were
also considered recently. Novel ideas for spin-boson models simulated with trapped
ions can be found in Ref.~\cite{2017arXiv170400629L}.

In Ref.~\cite{HBHL}, it has been proposed to use trapped ions for solving difficult optimization problems via quantum annealing. Such scheme, applied to the concrete example of number-partitioning, has come under scrutiny in Ref.~\cite{ncomms11524}. The idea 
is to profit from the known mapping between the number partitioning problem and the 
ground state of spin Hamiltonians~\cite{0305-4470-15-10-028}. In the interesting domain,  where the number partitioning problem is notably 
difficult, the system is actually in the spin-glass-like phase, which renders 
finding the actual ground state an involved task for classical methods. The annealing 
method proposed was found to work well for small number of ions at zero temperature. 
In this work we explore in detail a semiclassical approximation to the original problem,  
where the quantum correlations between the spins and the phonon bath are neglected. 
This, however, allows us to solve the Heisenberg equations of motion in an efficient 
way for much larger ion systems. Notably, the approach allows us to explore finite 
temperature effects on the annealing protocol.   

Our present work is structured as follows.  In Sec.~\ref{sec:start} we explain the specific 
technical details of the studied model and the calculations. In particular, the 
Hamiltonian is explained in~\ref{sec:Ham}, the reduction methods in~\ref{sec:SCdecoupling} 
and the annealing protocol~\ref{sec:annprot} along with the basic definitions~\ref{sec:fide}.
In Sec.~\ref{secIII}, we introduce our semiclassical approximation, and benchmark it with a full quantum treatment.
In Sec.~\ref{sec:CalcsAtTfin} we show the results of the semiclassical approach applied to finite 
temperature. Here, we set the initial phonon population to non-zero thermal values. A summary 
and our main conclusions are provided in Sect.~\ref{sec:concl}. Finally, in the 
appendix we provide some further tests to our numerical integration 
method (appendix ~\ref{app:Integrationtests}), and a brief study of the optimal bias for 
the annealing protocol (appendix ~\ref{sec:optim}).

\section{System}

\label{sec:start}
\subsection{Hamiltonian}
\label{sec:Ham}

We study a chain of $N$ trapped ions interacting by effective spin-spin interactions 
subjected to a transverse time-dependent magnetic field. Interactions are generated 
by Raman coupling the pseudo-spin degrees of freedom to the phonon modes which 
are obtained expanding the Coulomb force between the ions around their equilibrium 
positions~\cite{PhysRevLett.92.207901}. The phonon spectrum is defined through its 
natural frequencies $\omega_k$ and modes $\xi^{i}_{k}$. The dynamics of the system 
is described by a time-dependent Hamiltonian that in the Schr\"odinger picture reads,
\beqa
{\cal H}_{\rm S} \left( t \right)/\hbar &= & 
\sum_{k}^{M} {\omega_k \hat{a}^{\dagger}_k \hat{a}_k}
+\sum_{i,k}^{N,M} {\Omega \eta^{(i)}_{k} 
\sin{\left(\omega_{\rm L} t\right)} \left(\hat{a}^{\dagger}_k 
+ \hat{a}_k\right) \sigma_{x}^{\left( i \right)}} \nonumber\\
&+ & \sum_{i}^{N} {B \left( t \right) 
\sigma_{z}^{\left( i \right)}} + \varepsilon \sigma_{x}^{\left( p \right)},
\label{eq:Hamiltonian}
\eeqa
where $\hat{a}_k$($ \hat{a}^{\dagger}_k$) is the annihilation (creation) operator 
of one phonon in the $k$-th mode and $\omega_k$ is the frequency of that mode.  
The operators $\sigma_{x}^{\left( i \right)}$, $\sigma_{y}^{\left( i \right)}$, and 
$\sigma_{z}^{\left( i \right)}$ are the spin operators in the $i$-th position. 
The frequencies $\Omega$ and $\omega_{\rm L}$ are the Rabi frequency and the 
beatnote frequency of the laser, respectively. The dimensionless parameters $\eta^{(i)}_{k} $ are
the Lamb-Dicke parameters proportional to the displacement of an ion $i$ in 
the vibrational mode $k$, see Ref.~\cite{PhysRevLett.92.207901}. As usual, $t$ is 
time, and a time-dependent magnetic field $B(t)$ allows us to perform the quantum 
annealing. A small bias term, proportional to $\varepsilon$, has been added in 
the $p$-th position to remove the ${\cal Z}_2$ degeneracy. The upper limit of the sum over the ions $i$ is $N$, the number of 
ions. The upper limit of the sum in modes $k$ is $M$, the number of modes. The 
total number of phonon modes is $3N$, but the Raman beam couples to only $N$ modes, 
selected by the wave vector difference of the lasers. 
At this point, we may keep our analysis general by making no assumption about the number $M$ of modes.
However, all phonons which are considered are assumed to be coupled to the spin in the same way.
Hereinafter, the upper limits of the sums will be omitted for brevity. 

\subsection{Equations of motion}
\label{sec:SCdecoupling}

We compute the Heisenberg equations of motion for the quantum average of every 
operator in the Hamiltonian, $\hat{a}_k$, $\hat{a}^{\dagger}_k$, $\sigma_{x}^{\left( i \right)}$, 
$\sigma_{y}^{\left( i \right)}$, and $\sigma_{z}^{\left( i \right)}$. Given that these are time 
independent operators, the calculation reduces to commutators. Additionally, we 
replace $\hat{a}^{\dagger}_k+\hat{a}_k$ by $2 \Re \left[ \hat{a}_k\right]$ and 
$\hat{a}_k-\hat{a}^{\dagger}_k$ by $2 {\rm i} \Im \left[ \hat{a}_k\right]$. The 
equations of motion (all with real coefficients) read,
\beqa
\frac{d \langle \Re \left[ \hat{a}_k\right] \rangle}{dt} &=& 
\omega_k \langle \Im \left[ \hat{a}_k\right] \rangle \label{eem}
\\
\frac{d \langle \Im \left[ \hat{a}_k\right] \rangle}{dt} &=& 
-\omega_k \langle \Re \left[ \hat{a}_k\right] \rangle-
\sin{\left(\omega_{\rm L} t\right)} \sum_{j} \Omega \eta^{(j)}_{k} 
\langle \sigma_{x}^{\left( j \right)} \rangle,
\nonumber\\
\frac{d \langle \sigma_{x}^{\left( i \right)} \rangle}{dt} &=& 
-2 B\left( t \right) \langle\sigma_{y}^{\left( i \right)}\rangle,
\nonumber\\
\frac{d \langle \sigma_{y}^{\left( i \right)} \rangle}{dt} &= & -4
\sum_{l} \Omega \eta^{(i)}_{l} \sin{\left(\omega_{\rm L} t\right)} 
\langle \Re{ \left[\hat{a}_l\right]} \sigma_{z}^{\left( i \right)}\rangle
\nonumber\\
&&+2 B\left( t \right) \langle \sigma_{x}^{\left( i \right)}\rangle  
-2 \varepsilon \langle \sigma_{z}^{\left( i \right)} \rangle \delta_{p,i}, 
\nonumber\\
\frac{d \langle \sigma_{z}^{\left( i \right)} \rangle}{dt} &=&4
\sum_{l} \Omega \eta^{(i)}_{l} \sin{\left(\omega_{\rm L} t\right)} 
\langle \Re{ \left[\hat{a}_l\right]} \sigma_{y}^{\left( i \right)}\rangle
+2\varepsilon \langle \sigma_{y}^{\left( i \right)} \rangle \delta_{p,i}\nonumber\,.
\eeqa

\subsection{Annealing protocol}
\label{sec:annprot}

The functional form and value of $\Omega$, $B \left( t \right)$ and $\varepsilon$ determine 
the annealing protocol. In these annealing schemes, the initial value of the transverse 
magnetic field $B \left( t=0 \right)$ must be sufficiently strong to initialize the system 
in the paramagnetic phase, that is, $B$ must be larger than the effective spin-spin interactions 
$J \sim \Omega^2 \omega_{\rm rec}/ (\delta \omega_{\rm rad} )$, where $\omega_{\rm rec}$ is the 
recoil energy of the photon-ion coupling, $\omega_{\rm rad}$ is the radial trap frequency, and $\delta$ the detuning from the nearest phonon mode. 
For typical values, e.g. $\Omega \sim \delta \sim 100$ kHz, $\omega_{\rm rec}\sim 25$ kHz, 
$\omega_{\rm rad} \sim 5$ MHz, we obtain effective interactions $J\sim 1$ kHz, so we need an 
initial field strength $B(0)\sim 10$ kHz. The annealing scheme proceeds by turning down the 
magnetic field according to some functional form in order to adiabatically achieve the ground 
state of the Hamiltonian of interest. Given the adiabatic theorem, for a closed system 
initiallized in the ground state, the final system is guaranteed 
to be in the ground state as long as the system is gapped along the annealing path, and the 
variation is slow enough. Generalization to open systems has been proposed in Ref.~\cite{PhysRevA.93.032118}.

\begin{figure}[t]
\includegraphics[width=0.5\textwidth]{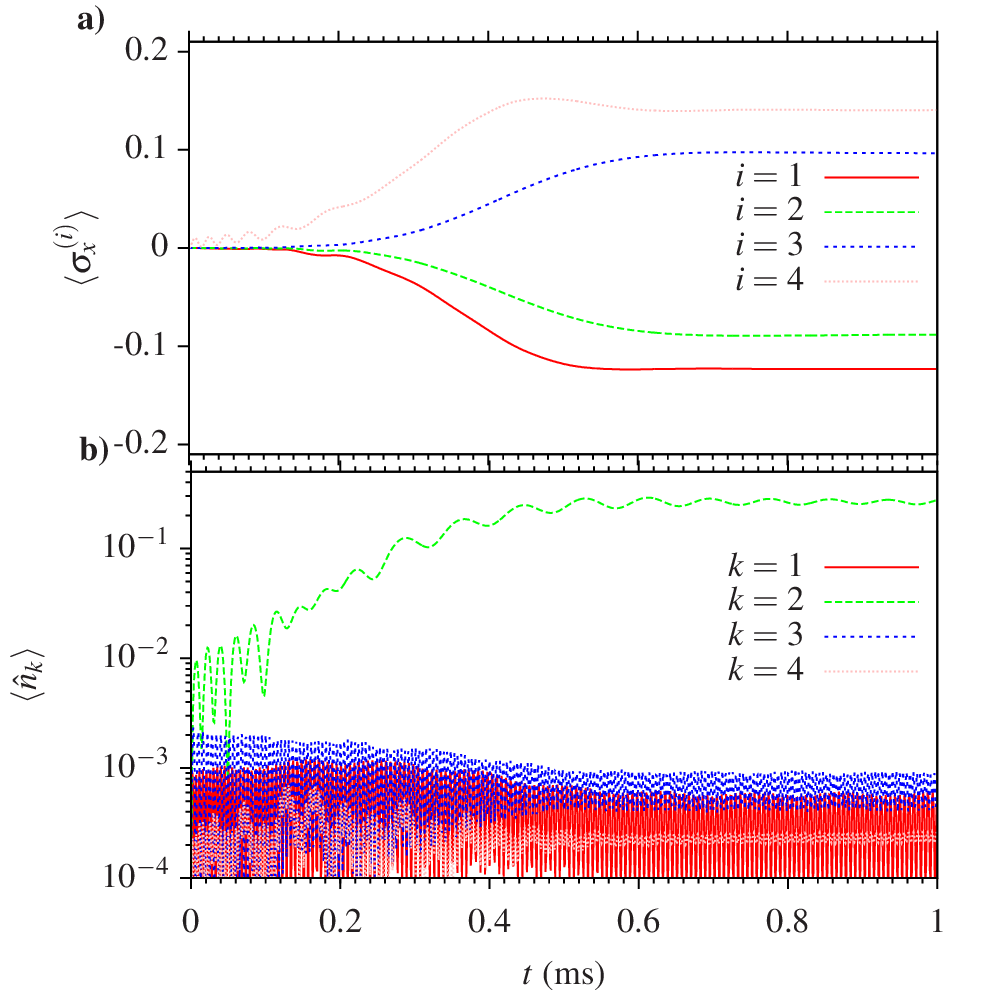}
\caption{(Upper panel) Evolution of $\expval{\hat{\sigma}_{x}^{\left( i \right)}}$ 
in a system with 4 spins, with initial populations of phononic modes set to $0$. 
The biased spin is the pink one, see text for details. (Lower panel) Evolution of 
the populations of the phononic modes in a system with 4 spins, with initial 
populations of phononic modes set to $0$. $\delta = 1$ MHz.}
\label{fig:sigma_x}
\end{figure}

We have used a decreasing exponential form for the transverse magnetic field, 
$B \left( t\right)=B(0)\,e^{-\frac{t}{\tau}}$ with a decay rate $\tau$. The other parameters, 
$\Omega$ and $\varepsilon$, remain constant. An example of the evolution of 
the system under this protocol is shown in Fig.~\ref{fig:sigma_x}. Initially, 
$\expval{\hat{\sigma}_{x}^{\left( i \right)}}=0$ for all $i$, and the total phononic 
population is set to $0$. Within tens of microseconds the phononic modes are 
populated. Not surprisingly, the mode next to the resonance becomes the most populated one, 
with a population being orders of magnitude larger than the population of the other 
modes. In contrast to these rapid changes of the phonon state, the spin dynamics is much 
slower. The spin expectations $\expval{\hat{\sigma}_{x}^{\left( i \right)}}$ remain mostly 
clustered around zero for hundreds of microseconds. When $B \left( t\right) \simeq \varepsilon$, 
the values $\expval{\hat{\sigma}_{x}^{\left( i \right)}}$ start to deviate from zero, and some 
acquire positive values, while others become negative. Thus, the spin curves separate 
from each other, and we call the time at which this happens the {\it separation time}. At 
some point after the separation time, the spin curves saturate, that is, from then on
$\expval{\hat{\sigma}_{x}^{\left( i \right)}}$ remain constant in time. We define the {\it waiting time} 
as the time when all $\expval{\hat{\sigma}_{x}^{\left( i \right)}}$ have stopped varying. 
At the waiting time, the phononic populations stabilize around certain values, although their 
oscillations do never vanish. 

The quantum annealer produces final values of $\expval{\hat{\sigma}_{x}^{\left( i \right)}}$ 
which are not fully polarized , that is, $|\expval{\hat{\sigma}_{x}^{\left( i \right)}}|<1$.
Thus, the final state differs from the classical ground state of the target Hamiltonian, that is, 
the Hamiltonian in the absence of a transverse field. Thus, we take as 
readout of the annealing protocol the average spin values~\cite{PhysRevE.94.032105,PhysRevA.96.042310,srep22318}. 
This is not a problem, as long as 
for all spins the sign matches with the one in the classical state. As explained in detail in Ref.~\cite{ncomms11524}, 
the spin configuration of the target Hamiltonian is determined by the dominant mode, 
defined as the one with frequency just below the beatnote frequency $\omega_{\rm L}$. 
There are different reasons why the final ground state might show a different spin 
pattern: Either, the annealing was too fast, that is, the value of $\tau$ was chosen too 
small, or the effective spin model is not valid. This is the case when  $\omega_{\rm L}$ is 
too close to a resonance $\omega_k$.

\subsection{Fidelity of the annealing protocol}
\label{sec:fide}

In order to quantify the success of the annealing protocol, that is, the 
ability of the method to identify the target ground state of the spin system, 
we will define the following fidelity, 
\begin{equation}
F=\begin{cases} 
      \min\limits_{i} {\left| \expval{\sigma_{x}^{(i)}}\right|} &{\rm if\;}
{\rm sign}{\left[\expval{\sigma_{x}^{(i)}}\right]}={\rm sign}{\left[ \eta_{k_d}^{(i)} \right]}, \forall i\\
		  0 & {\rm Otherwise}
   \end{cases},
\label{eq:Fide_def}
\end{equation}
where $\eta_{i j}$ is the $i$th component of the dominant mode $k_d$ for a fixed value 
of the beatnote frequency $\omega_{\rm L}$. That is, the fidelity is zero if the 
signs of $\langle \sigma_x^i\rangle$ do not match the signs of $\eta_i$ of the dominant 
mode. If the signs are reproduced, the value of the fidelity is defined as the smallest 
expectation value of the spins of the ions. Note that with this definition, any non-zero fidelity is good enough for correctly
identifying the ground state pattern, assuming the absence of noise in the system.

\section{Semiclassical approximation}
\label{secIII}
\subsection{Semiclassical equations of motion}

Now we will develop a semiclassical approximation to the exact equations of 
motion, Eq.~(\ref{eem}), that will allow us to study larger systems of ions 
and the effects of temperature on the annealing protocols. 
We make the following approximations:
\beqa
\langle \hat{a}_k \sigma_{\mu}^{\left( i \right)} \rangle &\simeq& 
\langle \hat{a}_k \rangle \langle \sigma_{\mu}^{\left( i \right)} \rangle \nonumber\\
\langle \hat{a}^{\dagger}_k \sigma_{\mu}^{\left( i \right)} \rangle &\simeq& \langle \hat{a}^{\dagger}_k \rangle \langle \sigma_{\mu}^{\left( i \right)} \rangle
\eeqa
with $\mu = \left\{ x,y,z\right\}$. These approximations ignore the quantum 
correlations in the coupling between bosonic and spin modes.

Additionally defining the auxiliary variables 
$S_k \left( t \right) \equiv \sin{\left(\omega_{\rm L} t\right)} 
\sum_{j} \Omega \eta^{j}_{k} \langle \sigma_{x}^{\left( j \right)} \rangle$ and 
$J^{\left( i \right)} \left( t \right) \equiv \sin{\left(\omega_{\rm L} t\right)} 
\Omega \left(2\sum_{l} \eta^{i}_{l} \langle \Re{ \left[\hat{a}_l\right]} 
\rangle \right) +\varepsilon \delta_{p,i}$, we obtain the approximate equations of motion,
\beqa
\frac{d \langle \Re \left[ \hat{a}_k\right] \rangle}{dt} &=& 
\omega_k \langle \Im \left[ \hat{a}_k\right] \rangle,
\label{eq:Eqs_motion_simpl}\\
\frac{d \langle \Im \left[ \hat{a}_k\right] \rangle}{dt} &=& 
-\omega_k \langle \Re \left[ \hat{a}_k\right] \rangle-S_k \left( t \right),
\nonumber\\
\frac{d \langle \vec{\sigma}^{\left( i \right)} \rangle}{dt} &=& -2
    \begin{pmatrix}
    0 & B\left( t \right) & 0\\
    -B\left( t \right) & 0 & J^{\left( i \right)} \left( t \right)\\
    0 & -J^{\left( i \right)} \left( t \right) & 0
    \end{pmatrix}
		\cdot \langle \vec{\sigma}^{\left( i \right)} \rangle\nonumber
\eeqa
where a spin vector notation, 
$\langle \vec{\sigma}^{\left( i \right)} \rangle =\left( \langle \sigma_{x}^{\left( i \right)} \rangle,
\langle \sigma_{y}^{\left( i \right)} \rangle, \langle \sigma_{z}^{\left( i \right)} \rangle \right)$, 
has been used. This is a system of $2\times M+ 3\times N$ non-linear first-order ordinary 
differential equations. Hence, it is numerically solved with a first order, ordinary 
differential equation solver that uses the Gragg–Bulirsch–Stoer method, stepsize control and 
order selection, called ODEX~\cite{ODEXcite}.

\subsection{Comparison of semiclassical approximation to full quantum evolution}
\begin{figure}[t]
\includegraphics[width=0.5\textwidth]{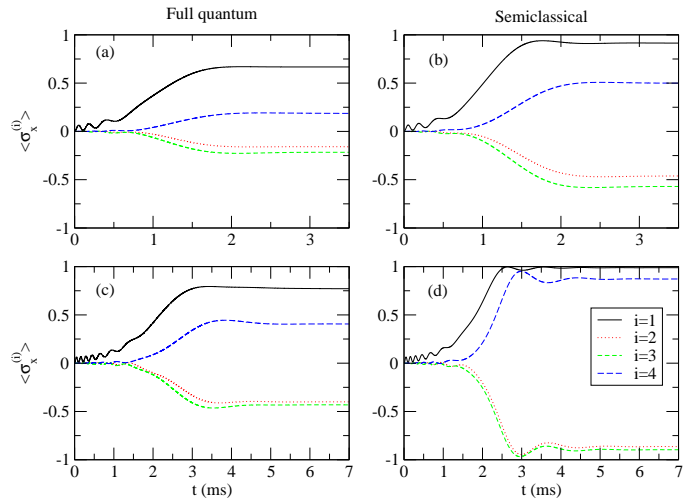}
\caption{\label{fig:fig4} Quantum (a,c) and semiclassical (b,d) evolution 
of $\expval{\hat{\sigma}_{x}^{\left( i \right)}}$. The figure is computed with $N=4$ ions 
with a detuning $\delta =2900$ kHz. Panels (a) and (b) correspond to $\tau=0.35$ ms 
and panels (c) and (d) to $\tau=0.7$ ms. }
\end{figure}

To benchmark the semiclassical method, we have compared it against a full quantum 
evolution of the system using Krylov subspaces. The latter is a method to study 
the dynamical evolution under time-dependent Hamiltonians that computes a reduced 
evolution operator omitting contributions smaller than a certain threshold, that is, 
transitions to irrelevant states. Despite this neglection, the Krylov evolution can be 
considered an \textit{exact} numerical simulation, as it iteratively determines which part of the 
Hilbert space is irrelevant at a given accuracy. 

In both cases, semiclassical and full quantum, the evolution of a given initial state 
under a time-depending Hamiltonian is calculated with time steps in a recurrent way. 
In the fully quantum calculation, the time steps are of the order of $1$ ns, while in 
the semiclassical description they are variable, but can  be orders of magnitude larger. 
In the quantum case, we have to specify a quantum state ---a complex vector in 
the joint Fock basis of phononic and spin modes---, containing the amplitudes of every 
state of the basis. In the semiclassical case, we only have to supply the initial mean 
values of every operator. 

It should be noted that the exact quantum evolution requires truncating the maximum 
phonon number which in our case was set to two phonons per mode. Such truncation of 
the Hilbert space requires sufficiently cool systems. And even with this truncation, 
the quantum evolution is restricted to a small number of ions. Considering only one 
transverse phonon branch, i.e. $N$ phonon modes, with a maximum population of two phonons 
per mode, the Hilbert space dimension is $2^N \times 3^N$, that is, a dimension of 46656 
for $N=6$ ions. The semiclassical approach, in contrast, allows us to explore larger systems. 

\begin{figure}[t]
\includegraphics[width=0.5\textwidth]{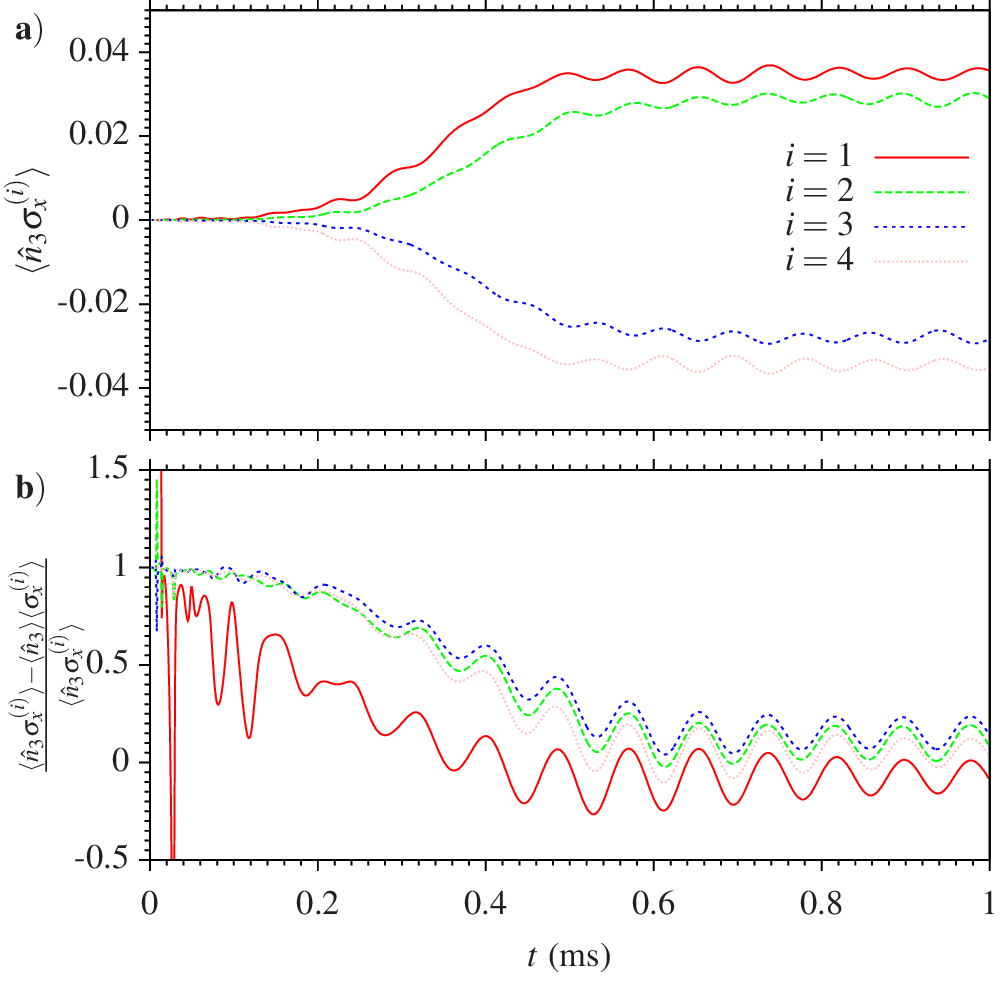}
\caption{Evolution of $\expval{\hat{n}_3 \hat{\sigma}_{x}^{\left( i \right)}}$ (a) and of 
the relative difference 
$(\expval{\hat{n}_3 \hat{\sigma}_{x}^{\left( i \right)}}-\expval{\hat{n}_3} \expval{\hat{\sigma}_{x}^{\left( i \right)}})/\expval{\hat{n}_3 \hat{\sigma}_{x}^{\left( i \right)}} $ 
for a system of four spins with initial populations of phonons set to zero for a 
detuning $\delta=1000$ kHz.
}
    \label{fig:diffs}
\end{figure}

\subsubsection{Time evolution of $\expval{\hat{\sigma}_{x}^{\left( i \right)}}$}

The semiclassical model captures well the qualitative behavior of the 
evolution of $\expval{\hat{\sigma}_{x}^{\left( i \right)}}$, as exemplified 
in Fig.~\ref{fig:fig4}. In the figure we compare the semiclassical evolution 
with the exact dynamics for $N=4$ ions for two different decay times. The 
discrepancy between semiclassical and exact evolution is smallest for shorter 
times ($t\lesssim \tau$), where the semiclassical model is able to correctly 
capture the details of the dynamical evolution, most notably little wiggles 
in the evolution of the biased ion, $i=1$. Importantly, the semiclassical model 
also agrees with the exact evolution regarding general features such as the 
separation time, and the sign of each $\expval{\hat{\sigma}_{x}^{\left( i \right)}}$ 
in the long-time limit. As discussed in more detail in the next paragraph, this 
enables a quite accurate prediction of annealing fidelities, despite the fact 
that the approximation disregards some quantum properties. Thus it provides 
a computationally efficient way to study the behavior of larger systems. 

For a better understanding of the errors in the semiclassical approach, we have 
exactly calculated the evolution of $\langle \hat n_3 \sigma_x^{(i)} \rangle$, see 
Fig.~\ref{fig:diffs}(a),  and of $\langle \hat n_3 \rangle \langle \sigma_x^{(i)} \rangle$. 
As our semiclassical approximation is based on substituting the former correlator 
by the latter one, the discrepancy between both expressions is an indicator for 
the quality of the semiclassical approach. In Fig.~\ref{fig:diffs}(b), we plot the 
relative difference as a function of time: Initially, the phonon and spin degrees of 
freedom are taken as uncorrelated, thus, the semiclassical and the exact description 
coincide at $t=0$. On short time scales, both correlators have small absolute values, 
but their relative difference becomes large. For times larger than the separation time, 
the absolute values of the correlators increase, and the relative errors decrease. 
On long time scales, the errors oscillate around mean values of the order 0.1.  This 
observation suggests that the main errors made in the semiclassical approximation are 
introduced at short times, where the transverse magnetic field and its temporal 
derivative takes large values. 

\begin{figure}[t]
\includegraphics[width=0.5\textwidth]{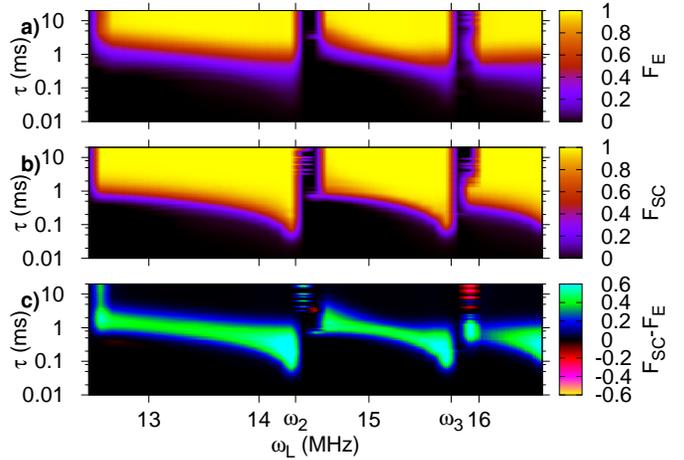}
\caption{\label{fig:QFide_na4} Fidelity obtained with the full quantum evolution 
$F_{\rm E}$ (a), the semiclassical model $F_{\rm SC}$ (b), and the difference between 
the semiclassical model and the full quantum (c). In both cases, we consider a system 
of four ions, and we plot the magnitudes as a function of $\omega_{\rm L}$ and 
$\tau$. The fidelity is readout after a time $20 \,\tau$.}
\end{figure}

\subsubsection{Fidelity}
\label{secfide}

As we have seen the semiclassical approximation provides a reasonable description 
of the dynamics in many configurations. Let us now explore in more detail in which 
parameter regions it predicts the correct fidelity for the annealing protocol. 
In Fig.~\ref{fig:QFide_na4} we present a comparison of the fidelities obtained from 
the exact time evolution and from the semiclassical approach for a system of four ions. 
We tune through a broad range of beatnote frequencies $\omega_L$, and vary the decay 
time $\tau$ of the magnetic field. The overall agreement is very good: Both methods 
predict a small fidelity when the field decays too fast (small $\tau$), or when the 
system is too close to one of the phonon resonances. The semiclassical evolution, however, 
slightly overestimates the fidelity for small $\tau$, and also slightly below each 
phonon resonance, that is, on the ferromagnetic side of the resonance. Notably, the 
semiclassical approach works quite well in the glassy regimes above the resonances, 
where it estimates correctly the regions in which the annealer fails for any annealing time.

As discussed earlier, the failure of the annealing protocol for small $\tau $ is due to 
non-adiabatic behavior in the fast varying field. The failure near the resonance, though, 
cannot be fixed by increasing $\tau$, and has its origin in the deviation from of the 
Dicke dynamics from the effective spin model. Although such deviations are expected on 
both sides of a phonon resonance, the region of zero fidelity is seen only on the glassy 
side of each resonance. From that perspective, the size of the spin gap seems to play a 
role as well, although in this regime we should not compare it to $\hbar/\tau$, but to 
those spin-phonon energy scales which are neglected in the effective spin model, that is, 
the first order term in a Magnus expansion, see Ref.~\cite{PhysRevLett.103.120502}.

\begin{figure}[t]
\includegraphics[width=0.5\textwidth]{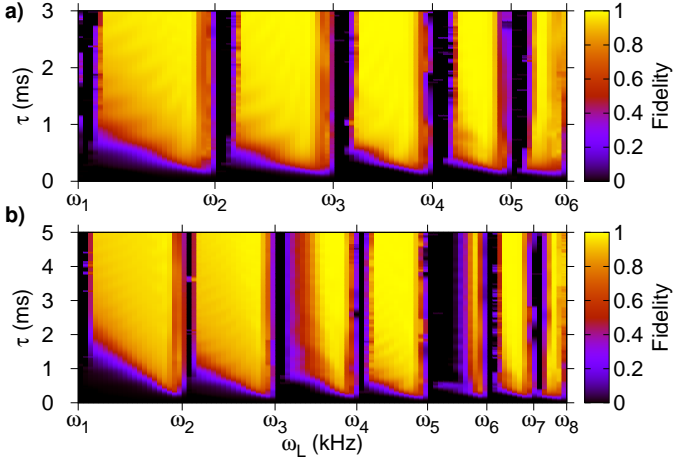}
\caption{\label{fig:SCFide_na6} Fidelity of a system of six ions (a) 
and a system of eight ions (b) as a function of  $\omega_{\rm L}$ and $\tau$ 
using the semiclassical approximation. Final times are $20 \tau$.}
\end{figure}

The main advantage of the semiclassical model is that it can easily be applied to larger 
systems. In Fig.~\ref{fig:SCFide_na6}, we consider systems of six and eight ions. Notably, 
a broad region of zero fidelity occurs for eight ions between $\omega_5$ and $\omega_6$. 
Its origin is unclear to us, and further calculations on the fully quantum evolution would 
be needed in order to discriminate whether they are true effects or merely calculation artifacts. 

We finish this section by discussing the factor which limits the scalability of the 
quantum annealer. As seen above, below a critical detuning from the resonance the 
fidelity drops to zero. This sets a limit to the scalability of the quantum annealer, 
because, with the number of modes being proportional to the number of ions, the mode 
spacing decreases when the system size is increased. However, we note that phonon 
spectra are not equidistant, and within the transverse branch, the phonon spacing 
is largest at the lower energetic end of the phonon spectrum. Thus, to achieve finite 
fidelity in an up-scaled system, one may need to operate in the regime of low-energy 
phonons. The semiclassical estimates in Ref.~\cite{ncomms11524} suggest that the 
quantum annealing still works in systems with more than twenty ions, presumably large 
enough to detect quantum speed-up.

\section{Finite temperature effects on the annealing protocol}
\label{sec:CalcsAtTfin}

Finite temperature effects are expected to reduce the quality of the annealing protocol, 
recently, however, thermal effects are shown to aid the annealing protocol 
for a 16-qubit problem in a superconducting setup~\cite{ncomms2920}. 
In this section we study the robustness of the annealing protocol when 
the phonons are initially at finite temperature. To do so, we consider an 
initial state with phonon mode populations set as follows: we fix the 
temperature $T$ of the phonons, then, the mean values of the number of phonons 
for each mode are sampled according to the a bosonic thermal bath probability 
distribution at $T$. With these initial conditions, the system is then evolved 
semiclassically according to Eqs.~(\ref{eq:Eqs_motion_simpl}). This process 
is repeated with different initial values of the population of phonons, sampled 
appropriately. After the evolution, the statistical moments are calculated in 
order to infer the thermal properties of the system at the final time. 

It should be noted that our dynamical model only captures the coherent Hamiltonian 
evolution, but no decoherence processes due to interactions with the environment. 
Thus, in order to account for all thermal effects, heating events, as they occur 
for instance due to trap inhomogeneties, should be taken into account by 
considering an increased initial temperature.

\subsection{Classical thermal phonons}

We assume that the initial populations of the phononic modes are determined by 
a phonon temperature. In the canonical ensemble, the expected value of the number 
operator of the phonons in the $k$-th mode is,
\begin{equation}
\langle \hat{n}_{k} \rangle = \frac{1}{e^{\beta \hbar \omega_k}-1}\,. 
\label{eq:n_k}
\end{equation}
The corresponding Hamiltonian of the symmetrized phonon field is 
$\hat{\cal H}_{\rm ph} = \hbar \sum_{k} \omega_k\frac{ \hat{a}^{\dagger}_k \hat{a}_k 
+ \hat{a}_k \hat{a}^{\dagger}_k}{2}$. The expected value of the annihilation 
operator $\alpha_k \equiv \expval{\hat{a}_k}$ 
is sampled as,
\begin{equation}
P\left(\alpha_k \right) = \frac{1}{\pi \langle \hat{n}_{k} \rangle} e^{-\frac{|\alpha_k|^2}{\langle \hat{n}_{k} \rangle}} \,.
\label{eq:p_alpha_k}
\end{equation}
The complex-valued Gaussian probability distribution function (PDF) 
for the random variable $\alpha_k$ is a product of two Normal PDFs ---one real 
the other purely imaginary---for the random variables $\Re\left[\alpha_k\right]$ 
and $\Im\left[\alpha_k\right]$. Both distributions have a mean $\mu=0$ and variance 
$\sigma^2=\hat{n}_{k}/2$. We thus use for convenience, 
\begin{equation}
P\left(\alpha_k \right) =
N \left(\Re\left[\alpha_k\right];0,\langle \hat{n}_{k} \rangle/2\right) 
N \left(\Im\left[\alpha_k\right];0,\langle \hat{n}_{k} \rangle/2\right)\,,
\label{eq:p_reimalpha_k}
\end{equation}
being $N \left(x;\mu,\sigma^2\right)$ the Normal PDF of the random variable $x$ with 
mean $\mu$ and variance $\sigma^2$.

\subsection{Effects of temperature on the protocol}

\begin{figure}[t!]
\includegraphics[width=0.5\textwidth]{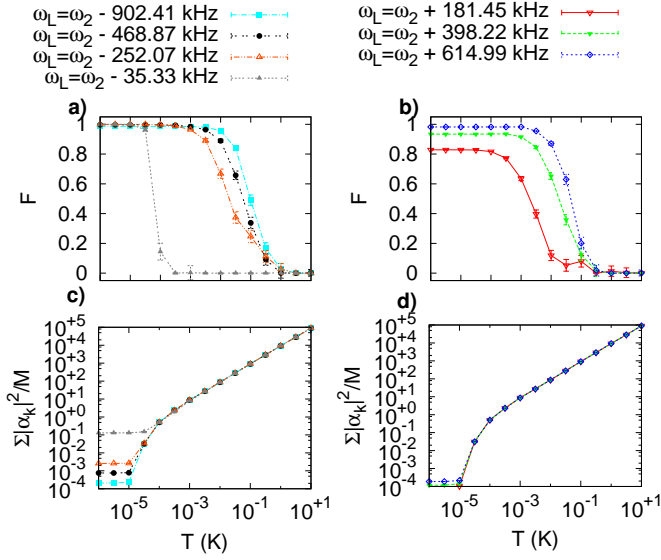}
\caption{\label{fig:ThermalFideandTotPop} 
We plot the fidelity (a,b) and the square of the phonon coherences (c,d), as a function 
of the initial phonon temperature, as obtained from the semiclassical calculation for 
a system of four ions. Different lines correspond to different detunings. Panels (a) and 
(c) consider cases where the second phonon resonance is approached from below (where 
the magnetic order is ferromagnetic), while panels (b) and (d) consider cases where 
the same resonance is approached from above (where a glassy regime occurs near the 
resonance). In our system, the second resonance occurs at a frequency $\omega_2=14332.7$ 
kHz. All calculations were done for $\tau = 10$ ms. Each point is obtained by sampling 
over 1,000 runs. The adscribed error is 2 $\sigma$.}
\end{figure}

To evaluate the effects of temperature on the proposed annealing protocol we will consider 
different initial temperatures and detunings. In all cases we will fix the decay time 
$\tau=10$ ms. Fig.~\ref{fig:ThermalFideandTotPop} shows the fidelity and the total phononic 
population per mode in the system. In the figure we compare results obtained with several 
values of $\omega_L$ and a broad range of temperatures of the phonons. Quite generally, panels 
(a) and (b) show that up to a certain temperature the fidelity is not affected by thermal 
phonons, but above this temperature the fidelity drops to zero. The value of this temperature 
strongly depends on the detuning, and decreases by several orders of magnitudes when we change 
from a far-detuned configuration to a near-resonance scenario. For instance, in the far-detuned 
regime at $\omega_{\rm L}=\omega_2-902.41$ kHz (solid squares) the critical temperature is of 
the order $0.1$ K, while close to the resonance at $\omega_{\rm L}=\omega_2-35.33$ kHz 
(small triangles), the fidelity drop occurs at a temperature of the order $10^{-4}$ K. Such 
behavior is seen both when we approach the phonon resonance from above (panel (b))or below 
(panel (a)), but we remind that above the resonance there is a finite region in which the 
fidelity is zero even at $T=0$, cf. the discussion of Fig.(~\ref{fig:QFide_na4}) in Sec. ~\ref{secfide}.

Regarding the phonon population, we assume that a reasonable estimate of $\langle \hat{n}_k \rangle$ 
is given by $|\alpha_k|^2$, as would be the case if the phonon field remains coherent. 
In Fig.~\ref{fig:ThermalFideandTotPop}(c,d), we may distinguish two different regimes: 
At low temperatures, i.e. for $T\lesssim 10^{-5}$ K, the final phonon population is dominated 
by those phonons which are produced by the spin-phonon coupling. The number of those phonons is 
independent from the temperature, and as shown earlier in Fig. 1(b), such phonons are generated 
also at $T=0$. The proximity to the resonance induces a larger population of the dominant mode, 
resulting in a larger value for the population as we approach the resonance. In contrast, 
for high temperatures, the phonon population is dominated by thermal phonons present already 
in the beginning of the evolution. In this case, the phonon population is more or less constant 
during the evolution, and the population number strongly depends on the temperature, according 
to the initial values from the Boltzmann distribution. As in this case, the time evolution does 
not noticeably change the phonon distribution, initial and final distribution are very close, 
and so are initial and final temperature. Between the low- and high- temperature phase, there 
is a narrow crossover regime, where the number of dynamically generated phonons is similar 
to the number of thermal phonons. The temperature at which this happens generally depends on 
the detuning, i.e. on $\omega_{\rm L}$. In all cases, the fidelity drop occurs only in the 
high-temperature regime, that is, the number of thermal photons must be large compared to the 
dynamically generated phonons in order to negatively affect the spin evolution. 

\subsection{Thermal tolerance of the Lamb-Dicke regime}

The Hamiltonian Eq.~(\ref{eq:Hamiltonian}) describes the trapped ion system when it is in 
the Lamb-Dicke regime, that is, for $k x \ll 1$. In a harmonic oscillator with frequency $\omega$, 
we have$\langle x^2 \rangle = \frac{\hbar}{m \omega} \left(\langle n \rangle + \frac{1}{2} \right)$.  
With $k=\sqrt{2m\omega_{\rm rec}/\hbar}$, and introducing the Lamb-Dicke parameter 
$\eta=\sqrt{\omega_{\rm rec}/\omega}$, we re-write the Lamb-Dicke condition as 
\begin{eqnarray}
\eta \sqrt{ 2\langle n \rangle +1 } \ll 1 \,.
\end{eqnarray}
All phonon frequencies $\omega$ are of the order of the radial trap frequency, 
$\omega_{\rm rad}=2\pi \times 2,655$ kHz. In our simulation, the recoil frequency is 
taken as $\omega_{\rm rec}=2\pi \times 15$ kHz, so we obtain a Lamb-Dicke parameter 
$\eta \approx 0.075$.  Thus, the Lamb-Dicke regime requires $\sqrt{\langle n \rangle +0.5} \ll 10$, 
which is fulfilled by a phonon occupation $\langle n \rangle \lesssim 1$. From that 
perspective, we have to disregard those calculations where the phonon population 
exceeds this number. From Fig.~\ref{fig:ThermalFideandTotPop}(c,d) we find that, for the 
Rabi frequency we have used, phonon numbers above 1 occur in the thermally dominated regime, 
independent from the detuning. This regime is characterized by temperatures $>10^{-4}$ K.

\section{Summary and conclusions}
\label{sec:concl}

We have considered a chain of trapped ions with an internal state (``spin'') coupled 
to vibrational modes via Raman lasers. The couplings are such that the effective model 
describing the ions is a long range spin model with tunable, pseudo-random couplings, 
leading to a spin-glass-like phase. The goal of our approach is the adiabatic distillation 
of the ground state in the glassy phase starting from a completely paramagnetic state. 
To this aim we consider the addition of a time-dependent transverse magnetic field. Our 
procedure goes as follows: At the initial time, the magnetic field is strong enough to 
ensure the ground state of the spins is a ferromagnetic state, with all spins aligned in 
the transverse direction. As time evolves we slowly, ideally adiabatically, remove the 
magnetic field such that the final Hamiltonian is our effective long-range spin model in the 
spin-glass-like phase.

We have simulated our annealing protocol using the exact evolution by means of a Krylov subspace method 
which is feasible for a small number of ions. In order to consider larger systems as well as to 
study the effect of temperature on the time evolution we have developed a semiclassical formalism which 
ignores the quantum correlations between the ions and the phonons. The quality of this method has been 
benchmarked by comparing its predictions with the exact evolution for four ions. The semiclassical 
model is found to provide a very accurate qualitative picture of our proposed method, and 
allows us to correctly identify the parameter region where the annealing protocol works well. 
By means of the semiclassical model we have thus extended our study to larger number of ions, providing 
an accurate picture of the ability of the annealing protocol to find the correct ground state depending 
on the annealing time. 

Finally, the semiclassical model has allowed us to study the robustness of the scheme 
for initial phonon states at finite temperature. We find that the effect of temperature 
strongly depends on the detuning from a phonon resonance. While in most configurations, 
the quantum annealing does not break down within the Lamb-Dicke regime, close to a resonance the situation is different. 
Here, the fidelity of the annealing may drop even before the Lamb-Dicke limit is reached, see 
Fig.~\ref{fig:ThermalFideandTotPop}.  Thus, while state-of-art spin model simulations which 
are carried out far off any phonon resonance (e.g. Refs. \cite{PhysRevLett.119.080501,Zhang2017}) 
require only cooling to the Lamb-Dicke limit, quantum annealing in the interesting glassy regime 
requires more cooling. Accordingly, our finding motivates the development of new, more efficient cooling 
techniques, as for instance cooling based on electromagnetically-induced 
transparency~\cite{PhysRevLett.85.5547,PhysRevLett.110.153002,PhysRevA.93.053401}, which is very well 
suited to simultaneously achieve low populations in all radial modes.

\begin{acknowledgments}
This work has been funded by EU grants (EQuaM (FP7-ICT-2013-C No. 323714), OSYRIS (ERC-2013-AdG No. 339106), 
SIQS (FP7-ICT-2011-9 No.  600645), and QUIC (H2020-FETPROACT-2014 No.  641122)), Spanish Ministerio de 
Econom\'ia y Competitividad grants (Severo Ochoa (SEV-2015-0522), FOQUS (FIS2013-46768-P), FIS2014-54672-P and 
FISICATEAMO (FIS2016-79508-P)), Generalitat de Catalunya (2014 SGR 401, 2014 SGR 874, and CERCA program), 
and Fundaci\'o Cellex.
TG acknowledges support of the NSF through the PFC@JQI.
\end{acknowledgments}

\bibliography{bib}

\begin{thebibliography}{51}%
\makeatletter
\providecommand \@ifxundefined [1]{%
 \@ifx{#1\undefined}
}%
\providecommand \@ifnum [1]{%
 \ifnum #1\expandafter \@firstoftwo
 \else \expandafter \@secondoftwo
 \fi
}%
\providecommand \@ifx [1]{%
 \ifx #1\expandafter \@firstoftwo
 \else \expandafter \@secondoftwo
 \fi
}%
\providecommand \natexlab [1]{#1}%
\providecommand \enquote  [1]{``#1''}%
\providecommand \bibnamefont  [1]{#1}%
\providecommand \bibfnamefont [1]{#1}%
\providecommand \citenamefont [1]{#1}%
\providecommand \href@noop [0]{\@secondoftwo}%
\providecommand \href [0]{\begingroup \@sanitize@url \@href}%
\providecommand \@href[1]{\@@startlink{#1}\@@href}%
\providecommand \@@href[1]{\endgroup#1\@@endlink}%
\providecommand \@sanitize@url [0]{\catcode `\\12\catcode `\$12\catcode
  `\&12\catcode `\#12\catcode `\^12\catcode `\_12\catcode `\%12\relax}%
\providecommand \@@startlink[1]{}%
\providecommand \@@endlink[0]{}%
\providecommand \url  [0]{\begingroup\@sanitize@url \@url }%
\providecommand \@url [1]{\endgroup\@href {#1}{\urlprefix }}%
\providecommand \urlprefix  [0]{URL }%
\providecommand \Eprint [0]{\href }%
\providecommand \doibase [0]{http://dx.doi.org/}%
\providecommand \selectlanguage [0]{\@gobble}%
\providecommand \bibinfo  [0]{\@secondoftwo}%
\providecommand \bibfield  [0]{\@secondoftwo}%
\providecommand \translation [1]{[#1]}%
\providecommand \BibitemOpen [0]{}%
\providecommand \bibitemStop [0]{}%
\providecommand \bibitemNoStop [0]{.\EOS\space}%
\providecommand \EOS [0]{\spacefactor3000\relax}%
\providecommand \BibitemShut  [1]{\csname bibitem#1\endcsname}%
\let\auto@bib@innerbib\@empty
%</preamble>
\bibitem [{\citenamefont {{Ac{\'i}n}}\ \emph {et~al.}(2017)\citenamefont
  {{Ac{\'i}n}}, \citenamefont {{Bloch}}, \citenamefont {{Buhrman}},
  \citenamefont {{Calarco}}, \citenamefont {{Eichler}}, \citenamefont
  {{Eisert}}, \citenamefont {{Esteve}}, \citenamefont {{Gisin}}, \citenamefont
  {{Glaser}}, \citenamefont {{Jelezko}}, \citenamefont {{Kuhr}}, \citenamefont
  {{Lewenstein}}, \citenamefont {{Riedel}}, \citenamefont {{Schmidt}},
  \citenamefont {{Thew}}, \citenamefont {{Wallraff}}, \citenamefont
  {{Walmsley}},\ and\ \citenamefont {{Wilhelm}}}]{2017arXiv171203773A}%
  \BibitemOpen
  \bibfield  {author} {\bibinfo {author} {\bibfnamefont {A.}~\bibnamefont
  {{Ac{\'i}n}}}, \bibinfo {author} {\bibfnamefont {I.}~\bibnamefont {{Bloch}}},
  \bibinfo {author} {\bibfnamefont {H.}~\bibnamefont {{Buhrman}}}, \bibinfo
  {author} {\bibfnamefont {T.}~\bibnamefont {{Calarco}}}, \bibinfo {author}
  {\bibfnamefont {C.}~\bibnamefont {{Eichler}}}, \bibinfo {author}
  {\bibfnamefont {J.}~\bibnamefont {{Eisert}}}, \bibinfo {author}
  {\bibfnamefont {D.}~\bibnamefont {{Esteve}}}, \bibinfo {author}
  {\bibfnamefont {N.}~\bibnamefont {{Gisin}}}, \bibinfo {author} {\bibfnamefont
  {S.~J.}\ \bibnamefont {{Glaser}}}, \bibinfo {author} {\bibfnamefont
  {F.}~\bibnamefont {{Jelezko}}}, \bibinfo {author} {\bibfnamefont
  {S.}~\bibnamefont {{Kuhr}}}, \bibinfo {author} {\bibfnamefont
  {M.}~\bibnamefont {{Lewenstein}}}, \bibinfo {author} {\bibfnamefont {M.~F.}\
  \bibnamefont {{Riedel}}}, \bibinfo {author} {\bibfnamefont {P.~O.}\
  \bibnamefont {{Schmidt}}}, \bibinfo {author} {\bibfnamefont {R.}~\bibnamefont
  {{Thew}}}, \bibinfo {author} {\bibfnamefont {A.}~\bibnamefont {{Wallraff}}},
  \bibinfo {author} {\bibfnamefont {I.}~\bibnamefont {{Walmsley}}}, \ and\
  \bibinfo {author} {\bibfnamefont {F.~K.}\ \bibnamefont {{Wilhelm}}},\
  }\href@noop {} {\bibfield  {journal} {\bibinfo  {journal} {ArXiv e-prints}\ }
  (\bibinfo {year} {2017})},\ \Eprint {http://arxiv.org/abs/1712.03773}
  {arXiv:1712.03773 [quant-ph]} \BibitemShut {NoStop}%
\bibitem [{\citenamefont {Barends}\ \emph {et~al.}(2014)\citenamefont
  {Barends}, \citenamefont {Kelly}, \citenamefont {Megrant}, \citenamefont
  {Veitia}, \citenamefont {Sank}, \citenamefont {Jeffrey}, \citenamefont
  {White}, \citenamefont {Mutus}, \citenamefont {Fowler}, \citenamefont
  {Campbell}, \citenamefont {Chen}, \citenamefont {Chen}, \citenamefont
  {Chiaro}, \citenamefont {Dunsworth}, \citenamefont {Neill}, \citenamefont
  {O'Malley}, \citenamefont {Roushan}, \citenamefont {Vainsencher},
  \citenamefont {Wenner}, \citenamefont {Korotkov}, \citenamefont {Cleland},\
  and\ \citenamefont {Martinis}}]{supercon}%
  \BibitemOpen
  \bibfield  {author} {\bibinfo {author} {\bibfnamefont {R.}~\bibnamefont
  {Barends}}, \bibinfo {author} {\bibfnamefont {J.}~\bibnamefont {Kelly}},
  \bibinfo {author} {\bibfnamefont {A.}~\bibnamefont {Megrant}}, \bibinfo
  {author} {\bibfnamefont {A.}~\bibnamefont {Veitia}}, \bibinfo {author}
  {\bibfnamefont {D.}~\bibnamefont {Sank}}, \bibinfo {author} {\bibfnamefont
  {E.}~\bibnamefont {Jeffrey}}, \bibinfo {author} {\bibfnamefont {T.~C.}\
  \bibnamefont {White}}, \bibinfo {author} {\bibfnamefont {J.}~\bibnamefont
  {Mutus}}, \bibinfo {author} {\bibfnamefont {A.~G.}\ \bibnamefont {Fowler}},
  \bibinfo {author} {\bibfnamefont {B.}~\bibnamefont {Campbell}}, \bibinfo
  {author} {\bibfnamefont {Y.}~\bibnamefont {Chen}}, \bibinfo {author}
  {\bibfnamefont {Z.}~\bibnamefont {Chen}}, \bibinfo {author} {\bibfnamefont
  {B.}~\bibnamefont {Chiaro}}, \bibinfo {author} {\bibfnamefont
  {A.}~\bibnamefont {Dunsworth}}, \bibinfo {author} {\bibfnamefont
  {C.}~\bibnamefont {Neill}}, \bibinfo {author} {\bibfnamefont
  {P.}~\bibnamefont {O'Malley}}, \bibinfo {author} {\bibfnamefont
  {P.}~\bibnamefont {Roushan}}, \bibinfo {author} {\bibfnamefont
  {A.}~\bibnamefont {Vainsencher}}, \bibinfo {author} {\bibfnamefont
  {J.}~\bibnamefont {Wenner}}, \bibinfo {author} {\bibfnamefont {A.~N.}\
  \bibnamefont {Korotkov}}, \bibinfo {author} {\bibfnamefont {A.~N.}\
  \bibnamefont {Cleland}}, \ and\ \bibinfo {author} {\bibfnamefont {J.~M.}\
  \bibnamefont {Martinis}},\ }\href {http://dx.doi.org/10.1038/nature13171}
  {\bibfield  {journal} {\bibinfo  {journal} {Nature}\ }\textbf {\bibinfo
  {volume} {508}},\ \bibinfo {pages} {500 EP} (\bibinfo {year}
  {2014})}\BibitemShut {NoStop}%
\bibitem [{\citenamefont {Linke}\ \emph {et~al.}(2017)\citenamefont {Linke},
  \citenamefont {Maslov}, \citenamefont {Roetteler}, \citenamefont {Debnath},
  \citenamefont {Figgatt}, \citenamefont {Landsman}, \citenamefont {Wright},\
  and\ \citenamefont {Monroe}}]{Linke3305}%
  \BibitemOpen
  \bibfield  {author} {\bibinfo {author} {\bibfnamefont {N.~M.}\ \bibnamefont
  {Linke}}, \bibinfo {author} {\bibfnamefont {D.}~\bibnamefont {Maslov}},
  \bibinfo {author} {\bibfnamefont {M.}~\bibnamefont {Roetteler}}, \bibinfo
  {author} {\bibfnamefont {S.}~\bibnamefont {Debnath}}, \bibinfo {author}
  {\bibfnamefont {C.}~\bibnamefont {Figgatt}}, \bibinfo {author} {\bibfnamefont
  {K.~A.}\ \bibnamefont {Landsman}}, \bibinfo {author} {\bibfnamefont
  {K.}~\bibnamefont {Wright}}, \ and\ \bibinfo {author} {\bibfnamefont
  {C.}~\bibnamefont {Monroe}},\ }\href {\doibase 10.1073/pnas.1618020114}
  {\bibfield  {journal} {\bibinfo  {journal} {Proceedings of the National
  Academy of Sciences}\ }\textbf {\bibinfo {volume} {114}},\ \bibinfo {pages}
  {3305} (\bibinfo {year} {2017})}\BibitemShut {NoStop}%
\bibitem [{\citenamefont {Monroe}\ and\ \citenamefont
  {Kim}(2013)}]{Monroe1164}%
  \BibitemOpen
  \bibfield  {author} {\bibinfo {author} {\bibfnamefont {C.}~\bibnamefont
  {Monroe}}\ and\ \bibinfo {author} {\bibfnamefont {J.}~\bibnamefont {Kim}},\
  }\href {\doibase 10.1126/science.1231298} {\bibfield  {journal} {\bibinfo
  {journal} {Science}\ }\textbf {\bibinfo {volume} {339}},\ \bibinfo {pages}
  {1164} (\bibinfo {year} {2013})}\BibitemShut {NoStop}%
\bibitem [{\citenamefont {Devoret}\ and\ \citenamefont
  {Schoelkopf}(2013)}]{Devoret1169}%
  \BibitemOpen
  \bibfield  {author} {\bibinfo {author} {\bibfnamefont {M.~H.}\ \bibnamefont
  {Devoret}}\ and\ \bibinfo {author} {\bibfnamefont {R.~J.}\ \bibnamefont
  {Schoelkopf}},\ }\href {\doibase 10.1126/science.1231930} {\bibfield
  {journal} {\bibinfo  {journal} {Science}\ }\textbf {\bibinfo {volume}
  {339}},\ \bibinfo {pages} {1169} (\bibinfo {year} {2013})}\BibitemShut
  {NoStop}%
\bibitem [{\citenamefont {Kirkpatrick}\ \emph {et~al.}(1983)\citenamefont
  {Kirkpatrick}, \citenamefont {Gelatt},\ and\ \citenamefont
  {Vecchi}}]{Kirkpatrick671}%
  \BibitemOpen
  \bibfield  {author} {\bibinfo {author} {\bibfnamefont {S.}~\bibnamefont
  {Kirkpatrick}}, \bibinfo {author} {\bibfnamefont {C.~D.}\ \bibnamefont
  {Gelatt}}, \ and\ \bibinfo {author} {\bibfnamefont {M.~P.}\ \bibnamefont
  {Vecchi}},\ }\href {\doibase 10.1126/science.220.4598.671} {\bibfield
  {journal} {\bibinfo  {journal} {Science}\ }\textbf {\bibinfo {volume}
  {220}},\ \bibinfo {pages} {671} (\bibinfo {year} {1983})}\BibitemShut
  {NoStop}%
\bibitem [{\citenamefont {\v{C}ern{\'y}}(1985)}]{cerny1985}%
  \BibitemOpen
  \bibfield  {author} {\bibinfo {author} {\bibfnamefont {V.}~\bibnamefont
  {\v{C}ern{\'y}}},\ }\href {\doibase 10.1007/BF00940812} {\bibfield  {journal}
  {\bibinfo  {journal} {Journal of Optimization Theory and Applications}\
  }\textbf {\bibinfo {volume} {45}},\ \bibinfo {pages} {41} (\bibinfo {year}
  {1985})}\BibitemShut {NoStop}%
\bibitem [{\citenamefont {Brooke}\ \emph {et~al.}(1999)\citenamefont {Brooke},
  \citenamefont {Bitko}, \citenamefont {Rosenbaum},\ and\ \citenamefont
  {Aeppli}}]{Brooke779}%
  \BibitemOpen
  \bibfield  {author} {\bibinfo {author} {\bibfnamefont {J.}~\bibnamefont
  {Brooke}}, \bibinfo {author} {\bibfnamefont {D.}~\bibnamefont {Bitko}},
  \bibinfo {author} {\bibfnamefont {T.~F.}\ \bibnamefont {Rosenbaum}}, \ and\
  \bibinfo {author} {\bibfnamefont {G.}~\bibnamefont {Aeppli}},\ }\href
  {\doibase 10.1126/science.284.5415.779} {\bibfield  {journal} {\bibinfo
  {journal} {Science}\ }\textbf {\bibinfo {volume} {284}},\ \bibinfo {pages}
  {779} (\bibinfo {year} {1999})},\ \Eprint
  {http://arxiv.org/abs/http://science.sciencemag.org/content/284/5415/779.full.pdf}
  {http://science.sciencemag.org/content/284/5415/779.full.pdf} \BibitemShut
  {NoStop}%
\bibitem [{\citenamefont {Kadowaki}\ and\ \citenamefont
  {Nishimori}(1998)}]{PhysRevE.58.5355}%
  \BibitemOpen
  \bibfield  {author} {\bibinfo {author} {\bibfnamefont {T.}~\bibnamefont
  {Kadowaki}}\ and\ \bibinfo {author} {\bibfnamefont {H.}~\bibnamefont
  {Nishimori}},\ }\href {\doibase 10.1103/PhysRevE.58.5355} {\bibfield
  {journal} {\bibinfo  {journal} {Phys. Rev. E}\ }\textbf {\bibinfo {volume}
  {58}},\ \bibinfo {pages} {5355} (\bibinfo {year} {1998})}\BibitemShut
  {NoStop}%
\bibitem [{\citenamefont {Farhi}\ \emph {et~al.}(2001)\citenamefont {Farhi},
  \citenamefont {Goldstone}, \citenamefont {Gutmann}, \citenamefont {Lapan},
  \citenamefont {Lundgren},\ and\ \citenamefont {Preda}}]{Farhi472}%
  \BibitemOpen
  \bibfield  {author} {\bibinfo {author} {\bibfnamefont {E.}~\bibnamefont
  {Farhi}}, \bibinfo {author} {\bibfnamefont {J.}~\bibnamefont {Goldstone}},
  \bibinfo {author} {\bibfnamefont {S.}~\bibnamefont {Gutmann}}, \bibinfo
  {author} {\bibfnamefont {J.}~\bibnamefont {Lapan}}, \bibinfo {author}
  {\bibfnamefont {A.}~\bibnamefont {Lundgren}}, \ and\ \bibinfo {author}
  {\bibfnamefont {D.}~\bibnamefont {Preda}},\ }\href {\doibase
  10.1126/science.1057726} {\bibfield  {journal} {\bibinfo  {journal}
  {Science}\ }\textbf {\bibinfo {volume} {292}},\ \bibinfo {pages} {472}
  (\bibinfo {year} {2001})},\ \Eprint
  {http://arxiv.org/abs/http://science.sciencemag.org/content/292/5516/472.full.pdf}
  {http://science.sciencemag.org/content/292/5516/472.full.pdf} \BibitemShut
  {NoStop}%
\bibitem [{\citenamefont {Albash}\ and\ \citenamefont
  {Lidar}(2018)}]{RevModPhys.90.015002}%
  \BibitemOpen
  \bibfield  {author} {\bibinfo {author} {\bibfnamefont {T.}~\bibnamefont
  {Albash}}\ and\ \bibinfo {author} {\bibfnamefont {D.~A.}\ \bibnamefont
  {Lidar}},\ }\href {\doibase 10.1103/RevModPhys.90.015002} {\bibfield
  {journal} {\bibinfo  {journal} {Rev. Mod. Phys.}\ }\textbf {\bibinfo {volume}
  {90}},\ \bibinfo {pages} {015002} (\bibinfo {year} {2018})}\BibitemShut
  {NoStop}%
\bibitem [{\citenamefont {Johnson}\ \emph {et~al.}(2011)\citenamefont
  {Johnson}, \citenamefont {Amin}, \citenamefont {Gildert}, \citenamefont
  {Lanting}, \citenamefont {Hamze}, \citenamefont {Dickson}, \citenamefont
  {Harris}, \citenamefont {Berkley}, \citenamefont {Johansson}, \citenamefont
  {Bunyk}, \citenamefont {Chapple}, \citenamefont {Enderud}, \citenamefont
  {Hilton}, \citenamefont {Karimi}, \citenamefont {Ladizinsky}, \citenamefont
  {Ladizinsky}, \citenamefont {Oh}, \citenamefont {Perminov}, \citenamefont
  {Rich}, \citenamefont {Thom}, \citenamefont {Tolkacheva}, \citenamefont
  {Truncik}, \citenamefont {Uchaikin}, \citenamefont {Wang}, \citenamefont
  {Wilson},\ and\ \citenamefont {Rose}}]{nature10012}%
  \BibitemOpen
  \bibfield  {author} {\bibinfo {author} {\bibfnamefont {M.~W.}\ \bibnamefont
  {Johnson}}, \bibinfo {author} {\bibfnamefont {M.~H.~S.}\ \bibnamefont
  {Amin}}, \bibinfo {author} {\bibfnamefont {S.}~\bibnamefont {Gildert}},
  \bibinfo {author} {\bibfnamefont {T.}~\bibnamefont {Lanting}}, \bibinfo
  {author} {\bibfnamefont {F.}~\bibnamefont {Hamze}}, \bibinfo {author}
  {\bibfnamefont {N.}~\bibnamefont {Dickson}}, \bibinfo {author} {\bibfnamefont
  {R.}~\bibnamefont {Harris}}, \bibinfo {author} {\bibfnamefont {A.~J.}\
  \bibnamefont {Berkley}}, \bibinfo {author} {\bibfnamefont {J.}~\bibnamefont
  {Johansson}}, \bibinfo {author} {\bibfnamefont {P.}~\bibnamefont {Bunyk}},
  \bibinfo {author} {\bibfnamefont {E.~M.}\ \bibnamefont {Chapple}}, \bibinfo
  {author} {\bibfnamefont {C.}~\bibnamefont {Enderud}}, \bibinfo {author}
  {\bibfnamefont {J.~P.}\ \bibnamefont {Hilton}}, \bibinfo {author}
  {\bibfnamefont {K.}~\bibnamefont {Karimi}}, \bibinfo {author} {\bibfnamefont
  {E.}~\bibnamefont {Ladizinsky}}, \bibinfo {author} {\bibfnamefont
  {N.}~\bibnamefont {Ladizinsky}}, \bibinfo {author} {\bibfnamefont
  {T.}~\bibnamefont {Oh}}, \bibinfo {author} {\bibfnamefont {I.}~\bibnamefont
  {Perminov}}, \bibinfo {author} {\bibfnamefont {C.}~\bibnamefont {Rich}},
  \bibinfo {author} {\bibfnamefont {M.~C.}\ \bibnamefont {Thom}}, \bibinfo
  {author} {\bibfnamefont {E.}~\bibnamefont {Tolkacheva}}, \bibinfo {author}
  {\bibfnamefont {C.~J.~S.}\ \bibnamefont {Truncik}}, \bibinfo {author}
  {\bibfnamefont {S.}~\bibnamefont {Uchaikin}}, \bibinfo {author}
  {\bibfnamefont {J.}~\bibnamefont {Wang}}, \bibinfo {author} {\bibfnamefont
  {B.}~\bibnamefont {Wilson}}, \ and\ \bibinfo {author} {\bibfnamefont
  {G.}~\bibnamefont {Rose}},\ }\href {\doibase 10.1038/nature10012} {\bibfield
  {journal} {\bibinfo  {journal} {Nature}\ }\textbf {\bibinfo {volume} {473}},\
  \bibinfo {pages} {194} (\bibinfo {year} {2011})}\BibitemShut {NoStop}%
\bibitem [{\citenamefont {R{\o}nnow}\ \emph {et~al.}(2014)\citenamefont
  {R{\o}nnow}, \citenamefont {Wang}, \citenamefont {Job}, \citenamefont
  {Boixo}, \citenamefont {Isakov}, \citenamefont {Wecker}, \citenamefont
  {Martinis}, \citenamefont {Lidar},\ and\ \citenamefont {Troyer}}]{Rnnow420}%
  \BibitemOpen
  \bibfield  {author} {\bibinfo {author} {\bibfnamefont {T.~F.}\ \bibnamefont
  {R{\o}nnow}}, \bibinfo {author} {\bibfnamefont {Z.}~\bibnamefont {Wang}},
  \bibinfo {author} {\bibfnamefont {J.}~\bibnamefont {Job}}, \bibinfo {author}
  {\bibfnamefont {S.}~\bibnamefont {Boixo}}, \bibinfo {author} {\bibfnamefont
  {S.~V.}\ \bibnamefont {Isakov}}, \bibinfo {author} {\bibfnamefont
  {D.}~\bibnamefont {Wecker}}, \bibinfo {author} {\bibfnamefont {J.~M.}\
  \bibnamefont {Martinis}}, \bibinfo {author} {\bibfnamefont {D.~A.}\
  \bibnamefont {Lidar}}, \ and\ \bibinfo {author} {\bibfnamefont
  {M.}~\bibnamefont {Troyer}},\ }\href {\doibase 10.1126/science.1252319}
  {\bibfield  {journal} {\bibinfo  {journal} {Science}\ }\textbf {\bibinfo
  {volume} {345}},\ \bibinfo {pages} {420} (\bibinfo {year}
  {2014})}\BibitemShut {NoStop}%
\bibitem [{\citenamefont {Boixo}\ \emph {et~al.}(2014)\citenamefont {Boixo},
  \citenamefont {R{\o}nnow}, \citenamefont {Isakov}, \citenamefont {Wang},
  \citenamefont {Wecker}, \citenamefont {Lidar}, \citenamefont {Martinis},\
  and\ \citenamefont {Troyer}}]{Boixo2014}%
  \BibitemOpen
  \bibfield  {author} {\bibinfo {author} {\bibfnamefont {S.}~\bibnamefont
  {Boixo}}, \bibinfo {author} {\bibfnamefont {T.~F.}\ \bibnamefont
  {R{\o}nnow}}, \bibinfo {author} {\bibfnamefont {S.~V.}\ \bibnamefont
  {Isakov}}, \bibinfo {author} {\bibfnamefont {Z.}~\bibnamefont {Wang}},
  \bibinfo {author} {\bibfnamefont {D.}~\bibnamefont {Wecker}}, \bibinfo
  {author} {\bibfnamefont {D.~A.}\ \bibnamefont {Lidar}}, \bibinfo {author}
  {\bibfnamefont {J.~M.}\ \bibnamefont {Martinis}}, \ and\ \bibinfo {author}
  {\bibfnamefont {M.}~\bibnamefont {Troyer}},\ }\href
  {http://dx.doi.org/10.1038/nphys2900} {\bibfield  {journal} {\bibinfo
  {journal} {Nature Physics}\ }\textbf {\bibinfo {volume} {10}},\ \bibinfo
  {pages} {218 EP} (\bibinfo {year} {2014})},\ \bibinfo {note}
  {article}\BibitemShut {NoStop}%
\bibitem [{\citenamefont {Lanting}()}]{youtube}%
  \BibitemOpen
  \bibfield  {author} {\bibinfo {author} {\bibfnamefont {T.~M.}\ \bibnamefont
  {Lanting}},\ }\href@noop {} {\enquote {\bibinfo {title} {D-wave 2000q},}\
  }\bibinfo {note} {Adiabatic Quantum Computing Conference 2017 (AQC
  2017)}\BibitemShut {NoStop}%
\bibitem [{\citenamefont {Mintert}\ and\ \citenamefont
  {Wunderlich}(2001)}]{PhysRevLett.87.257904}%
  \BibitemOpen
  \bibfield  {author} {\bibinfo {author} {\bibfnamefont {F.}~\bibnamefont
  {Mintert}}\ and\ \bibinfo {author} {\bibfnamefont {C.}~\bibnamefont
  {Wunderlich}},\ }\href {\doibase 10.1103/PhysRevLett.87.257904} {\bibfield
  {journal} {\bibinfo  {journal} {Phys. Rev. Lett.}\ }\textbf {\bibinfo
  {volume} {87}},\ \bibinfo {pages} {257904} (\bibinfo {year}
  {2001})}\BibitemShut {NoStop}%
\bibitem [{\citenamefont {Porras}\ and\ \citenamefont
  {Cirac}(2004)}]{PhysRevLett.92.207901}%
  \BibitemOpen
  \bibfield  {author} {\bibinfo {author} {\bibfnamefont {D.}~\bibnamefont
  {Porras}}\ and\ \bibinfo {author} {\bibfnamefont {J.~I.}\ \bibnamefont
  {Cirac}},\ }\href {\doibase 10.1103/PhysRevLett.92.207901} {\bibfield
  {journal} {\bibinfo  {journal} {Phys. Rev. Lett.}\ }\textbf {\bibinfo
  {volume} {92}},\ \bibinfo {pages} {207901} (\bibinfo {year}
  {2004})}\BibitemShut {NoStop}%
\bibitem [{\citenamefont {Friedenauer}\ \emph {et~al.}(2008)\citenamefont
  {Friedenauer}, \citenamefont {Schmitz}, \citenamefont {Glueckert},
  \citenamefont {Porras},\ and\ \citenamefont {Schaetz}}]{friedenauer08}%
  \BibitemOpen
  \bibfield  {author} {\bibinfo {author} {\bibfnamefont {A.}~\bibnamefont
  {Friedenauer}}, \bibinfo {author} {\bibfnamefont {H.}~\bibnamefont
  {Schmitz}}, \bibinfo {author} {\bibfnamefont {J.~T.}\ \bibnamefont
  {Glueckert}}, \bibinfo {author} {\bibfnamefont {D.}~\bibnamefont {Porras}}, \
  and\ \bibinfo {author} {\bibfnamefont {T.}~\bibnamefont {Schaetz}},\ }\href
  {\doibase 10.1038/nphys1032} {\bibfield  {journal} {\bibinfo  {journal} {Nat.
  Phys.}\ }\textbf {\bibinfo {volume} {4}},\ \bibinfo {pages} {757} (\bibinfo
  {year} {2008})}\BibitemShut {NoStop}%
\bibitem [{\citenamefont {Mielenz}\ \emph {et~al.}(2016)\citenamefont
  {Mielenz}, \citenamefont {Kalis}, \citenamefont {Wittemer}, \citenamefont
  {Hakelberg}, \citenamefont {Warring}, \citenamefont {Schmied}, \citenamefont
  {Blain}, \citenamefont {Maunz}, \citenamefont {Moehring}, \citenamefont
  {Leibfried},\ and\ \citenamefont {Schaetz}}]{2015arXiv151203559M}%
  \BibitemOpen
  \bibfield  {author} {\bibinfo {author} {\bibfnamefont {M.}~\bibnamefont
  {Mielenz}}, \bibinfo {author} {\bibfnamefont {H.}~\bibnamefont {Kalis}},
  \bibinfo {author} {\bibfnamefont {M.}~\bibnamefont {Wittemer}}, \bibinfo
  {author} {\bibfnamefont {F.}~\bibnamefont {Hakelberg}}, \bibinfo {author}
  {\bibfnamefont {U.}~\bibnamefont {Warring}}, \bibinfo {author} {\bibfnamefont
  {R.}~\bibnamefont {Schmied}}, \bibinfo {author} {\bibfnamefont
  {M.}~\bibnamefont {Blain}}, \bibinfo {author} {\bibfnamefont
  {P.}~\bibnamefont {Maunz}}, \bibinfo {author} {\bibfnamefont {D.~L.}\
  \bibnamefont {Moehring}}, \bibinfo {author} {\bibfnamefont {D.}~\bibnamefont
  {Leibfried}}, \ and\ \bibinfo {author} {\bibfnamefont {T.}~\bibnamefont
  {Schaetz}},\ }\href {\doibase 10.1038/ncomms11839;; 10.1038/ncomms11839}
  {\bibfield  {journal} {\bibinfo  {journal} {Nat. Commun.}\ }\textbf {\bibinfo
  {volume} {7}},\ \bibinfo {pages} {ncomms11839} (\bibinfo {year}
  {2016})}\BibitemShut {NoStop}%
\bibitem [{\citenamefont {Kalis}\ \emph {et~al.}(2016)\citenamefont {Kalis},
  \citenamefont {Hakelberg}, \citenamefont {Wittemer}, \citenamefont {Mielenz},
  \citenamefont {Warring},\ and\ \citenamefont {Schaetz}}]{PhysRevA.94.023401}%
  \BibitemOpen
  \bibfield  {author} {\bibinfo {author} {\bibfnamefont {H.}~\bibnamefont
  {Kalis}}, \bibinfo {author} {\bibfnamefont {F.}~\bibnamefont {Hakelberg}},
  \bibinfo {author} {\bibfnamefont {M.}~\bibnamefont {Wittemer}}, \bibinfo
  {author} {\bibfnamefont {M.}~\bibnamefont {Mielenz}}, \bibinfo {author}
  {\bibfnamefont {U.}~\bibnamefont {Warring}}, \ and\ \bibinfo {author}
  {\bibfnamefont {T.}~\bibnamefont {Schaetz}},\ }\href {\doibase
  10.1103/PhysRevA.94.023401} {\bibfield  {journal} {\bibinfo  {journal} {Phys.
  Rev. A}\ }\textbf {\bibinfo {volume} {94}},\ \bibinfo {pages} {023401}
  (\bibinfo {year} {2016})}\BibitemShut {NoStop}%
\bibitem [{\citenamefont {{Lambrecht}}\ \emph {et~al.}(2017)\citenamefont
  {{Lambrecht}}, \citenamefont {{Schmidt}}, \citenamefont {{Weckesser}},
  \citenamefont {{Debatin}}, \citenamefont {{Karpa}},\ and\ \citenamefont
  {{Schaetz}}}]{2016arXiv160906429L}%
  \BibitemOpen
  \bibfield  {author} {\bibinfo {author} {\bibfnamefont {A.}~\bibnamefont
  {{Lambrecht}}}, \bibinfo {author} {\bibfnamefont {J.}~\bibnamefont
  {{Schmidt}}}, \bibinfo {author} {\bibfnamefont {P.}~\bibnamefont
  {{Weckesser}}}, \bibinfo {author} {\bibfnamefont {M.}~\bibnamefont
  {{Debatin}}}, \bibinfo {author} {\bibfnamefont {L.}~\bibnamefont {{Karpa}}},
  \ and\ \bibinfo {author} {\bibfnamefont {T.}~\bibnamefont {{Schaetz}}},\
  }\href {\doibase 10.1038/s41566-017-0030-2} {\bibfield  {journal} {\bibinfo
  {journal} {Nature Photonics}\ }\textbf {\bibinfo {volume} {11}},\ \bibinfo
  {pages} {704} (\bibinfo {year} {2017})}\BibitemShut {NoStop}%
\bibitem [{\citenamefont {Jurcevic}\ \emph {et~al.}(2017)\citenamefont
  {Jurcevic}, \citenamefont {Shen}, \citenamefont {Hauke}, \citenamefont
  {Maier}, \citenamefont {Brydges}, \citenamefont {Hempel}, \citenamefont
  {Lanyon}, \citenamefont {Heyl}, \citenamefont {Blatt},\ and\ \citenamefont
  {Roos}}]{PhysRevLett.119.080501}%
  \BibitemOpen
  \bibfield  {author} {\bibinfo {author} {\bibfnamefont {P.}~\bibnamefont
  {Jurcevic}}, \bibinfo {author} {\bibfnamefont {H.}~\bibnamefont {Shen}},
  \bibinfo {author} {\bibfnamefont {P.}~\bibnamefont {Hauke}}, \bibinfo
  {author} {\bibfnamefont {C.}~\bibnamefont {Maier}}, \bibinfo {author}
  {\bibfnamefont {T.}~\bibnamefont {Brydges}}, \bibinfo {author} {\bibfnamefont
  {C.}~\bibnamefont {Hempel}}, \bibinfo {author} {\bibfnamefont {B.~P.}\
  \bibnamefont {Lanyon}}, \bibinfo {author} {\bibfnamefont {M.}~\bibnamefont
  {Heyl}}, \bibinfo {author} {\bibfnamefont {R.}~\bibnamefont {Blatt}}, \ and\
  \bibinfo {author} {\bibfnamefont {C.~F.}\ \bibnamefont {Roos}},\ }\href
  {\doibase 10.1103/PhysRevLett.119.080501} {\bibfield  {journal} {\bibinfo
  {journal} {Phys. Rev. Lett.}\ }\textbf {\bibinfo {volume} {119}},\ \bibinfo
  {pages} {080501} (\bibinfo {year} {2017})}\BibitemShut {NoStop}%
\bibitem [{\citenamefont {Li}\ \emph {et~al.}(2017)\citenamefont {Li},
  \citenamefont {Urban}, \citenamefont {Noel}, \citenamefont {Chuang},
  \citenamefont {Xia}, \citenamefont {Ransford}, \citenamefont {Hemmerling},
  \citenamefont {Wang}, \citenamefont {Li}, \citenamefont {H{\"a}ffner},\ and\
  \citenamefont {Zhang}}]{PhysRevLett.118.053001}%
  \BibitemOpen
  \bibfield  {author} {\bibinfo {author} {\bibfnamefont {H.-K.}\ \bibnamefont
  {Li}}, \bibinfo {author} {\bibfnamefont {E.}~\bibnamefont {Urban}}, \bibinfo
  {author} {\bibfnamefont {C.}~\bibnamefont {Noel}}, \bibinfo {author}
  {\bibfnamefont {A.}~\bibnamefont {Chuang}}, \bibinfo {author} {\bibfnamefont
  {Y.}~\bibnamefont {Xia}}, \bibinfo {author} {\bibfnamefont {A.}~\bibnamefont
  {Ransford}}, \bibinfo {author} {\bibfnamefont {B.}~\bibnamefont
  {Hemmerling}}, \bibinfo {author} {\bibfnamefont {Y.}~\bibnamefont {Wang}},
  \bibinfo {author} {\bibfnamefont {T.}~\bibnamefont {Li}}, \bibinfo {author}
  {\bibfnamefont {H.}~\bibnamefont {H{\"a}ffner}}, \ and\ \bibinfo {author}
  {\bibfnamefont {X.}~\bibnamefont {Zhang}},\ }\href {\doibase
  10.1103/PhysRevLett.118.053001} {\bibfield  {journal} {\bibinfo  {journal}
  {Phys. Rev. Lett.}\ }\textbf {\bibinfo {volume} {118}},\ \bibinfo {pages}
  {053001} (\bibinfo {year} {2017})}\BibitemShut {NoStop}%
\bibitem [{\citenamefont {Zhang}\ \emph
  {et~al.}(2017{\natexlab{a}})\citenamefont {Zhang}, \citenamefont {Hess},
  \citenamefont {Kyprianidis}, \citenamefont {Becker}, \citenamefont {Lee},
  \citenamefont {Smith}, \citenamefont {Pagano}, \citenamefont {Potirniche},
  \citenamefont {Potter}, \citenamefont {Vishwanath}, \citenamefont {Yao},\
  and\ \citenamefont {Monroe}}]{timecrystal}%
  \BibitemOpen
  \bibfield  {author} {\bibinfo {author} {\bibfnamefont {J.}~\bibnamefont
  {Zhang}}, \bibinfo {author} {\bibfnamefont {P.~W.}\ \bibnamefont {Hess}},
  \bibinfo {author} {\bibfnamefont {A.}~\bibnamefont {Kyprianidis}}, \bibinfo
  {author} {\bibfnamefont {P.}~\bibnamefont {Becker}}, \bibinfo {author}
  {\bibfnamefont {A.}~\bibnamefont {Lee}}, \bibinfo {author} {\bibfnamefont
  {J.}~\bibnamefont {Smith}}, \bibinfo {author} {\bibfnamefont
  {G.}~\bibnamefont {Pagano}}, \bibinfo {author} {\bibfnamefont {I.-D.}\
  \bibnamefont {Potirniche}}, \bibinfo {author} {\bibfnamefont {A.~C.}\
  \bibnamefont {Potter}}, \bibinfo {author} {\bibfnamefont {A.}~\bibnamefont
  {Vishwanath}}, \bibinfo {author} {\bibfnamefont {N.~Y.}\ \bibnamefont {Yao}},
  \ and\ \bibinfo {author} {\bibfnamefont {C.}~\bibnamefont {Monroe}},\ }\href
  {\doibase 10.1038/nature21413;;; 10.1038/nature21413} {\bibfield  {journal}
  {\bibinfo  {journal} {Nature}\ }\textbf {\bibinfo {volume} {543}},\ \bibinfo
  {pages} {217} (\bibinfo {year} {2017}{\natexlab{a}})}\BibitemShut {NoStop}%
\bibitem [{\citenamefont {Zhang}\ \emph
  {et~al.}(2017{\natexlab{b}})\citenamefont {Zhang}, \citenamefont {Pagano},
  \citenamefont {Hess}, \citenamefont {Kyprianidis}, \citenamefont {Becker},
  \citenamefont {Kaplan}, \citenamefont {Gorshkov}, \citenamefont {Gong},\ and\
  \citenamefont {Monroe}}]{Zhang2017}%
  \BibitemOpen
  \bibfield  {author} {\bibinfo {author} {\bibfnamefont {J.}~\bibnamefont
  {Zhang}}, \bibinfo {author} {\bibfnamefont {G.}~\bibnamefont {Pagano}},
  \bibinfo {author} {\bibfnamefont {P.~W.}\ \bibnamefont {Hess}}, \bibinfo
  {author} {\bibfnamefont {A.}~\bibnamefont {Kyprianidis}}, \bibinfo {author}
  {\bibfnamefont {P.}~\bibnamefont {Becker}}, \bibinfo {author} {\bibfnamefont
  {H.}~\bibnamefont {Kaplan}}, \bibinfo {author} {\bibfnamefont {A.~V.}\
  \bibnamefont {Gorshkov}}, \bibinfo {author} {\bibfnamefont {Z.-X.}\
  \bibnamefont {Gong}}, \ and\ \bibinfo {author} {\bibfnamefont
  {C.}~\bibnamefont {Monroe}},\ }\href {http://dx.doi.org/10.1038/nature24654}
  {\bibfield  {journal} {\bibinfo  {journal} {Nature}\ }\textbf {\bibinfo
  {volume} {551}},\ \bibinfo {pages} {601 EP} (\bibinfo {year}
  {2017}{\natexlab{b}})}\BibitemShut {NoStop}%
\bibitem [{\citenamefont {{Safavi-Naini}}\ \emph {et~al.}(2017)\citenamefont
  {{Safavi-Naini}}, \citenamefont {{Lewis-Swan}}, \citenamefont {{Bohnet}},
  \citenamefont {{Garttner}}, \citenamefont {{Gilmore}}, \citenamefont
  {{Jordan}}, \citenamefont {{Cohn}}, \citenamefont {{Freericks}},
  \citenamefont {{Rey}},\ and\ \citenamefont
  {{Bollinger}}}]{2017arXiv171107392S}%
  \BibitemOpen
  \bibfield  {author} {\bibinfo {author} {\bibfnamefont {A.}~\bibnamefont
  {{Safavi-Naini}}}, \bibinfo {author} {\bibfnamefont {R.~J.}\ \bibnamefont
  {{Lewis-Swan}}}, \bibinfo {author} {\bibfnamefont {J.~G.}\ \bibnamefont
  {{Bohnet}}}, \bibinfo {author} {\bibfnamefont {M.}~\bibnamefont
  {{Garttner}}}, \bibinfo {author} {\bibfnamefont {K.~A.}\ \bibnamefont
  {{Gilmore}}}, \bibinfo {author} {\bibfnamefont {E.}~\bibnamefont {{Jordan}}},
  \bibinfo {author} {\bibfnamefont {J.}~\bibnamefont {{Cohn}}}, \bibinfo
  {author} {\bibfnamefont {J.~K.}\ \bibnamefont {{Freericks}}}, \bibinfo
  {author} {\bibfnamefont {A.~M.}\ \bibnamefont {{Rey}}}, \ and\ \bibinfo
  {author} {\bibfnamefont {J.~J.}\ \bibnamefont {{Bollinger}}},\ }\href@noop {}
  {\bibfield  {journal} {\bibinfo  {journal} {ArXiv e-prints}\ } (\bibinfo
  {year} {2017})},\ \Eprint {http://arxiv.org/abs/1711.07392} {arXiv:1711.07392
  [quant-ph]} \BibitemShut {NoStop}%
\bibitem [{\citenamefont {Walls}\ and\ \citenamefont {Milburn}(2008)}]{TWA}%
  \BibitemOpen
  \bibfield  {author} {\bibinfo {author} {\bibfnamefont {D.}~\bibnamefont
  {Walls}}\ and\ \bibinfo {author} {\bibfnamefont {G.}~\bibnamefont
  {Milburn}},\ }\href@noop {} {\emph {\bibinfo {title} {{Quantum Optics}}}},\
  \bibinfo {edition} {2nd}\ ed.\ (\bibinfo  {publisher} {Springer},\ \bibinfo
  {year} {2008})\BibitemShut {NoStop}%
\bibitem [{\citenamefont {{Pi{\~n}eiro Orioli}}\ \emph
  {et~al.}(2017)\citenamefont {{Pi{\~n}eiro Orioli}}, \citenamefont
  {Safavi-Naini}, \citenamefont {Wall},\ and\ \citenamefont
  {Rey}}]{PhysRevA.96.033607}%
  \BibitemOpen
  \bibfield  {author} {\bibinfo {author} {\bibfnamefont {A.}~\bibnamefont
  {{Pi{\~n}eiro Orioli}}}, \bibinfo {author} {\bibfnamefont {A.}~\bibnamefont
  {Safavi-Naini}}, \bibinfo {author} {\bibfnamefont {M.~L.}\ \bibnamefont
  {Wall}}, \ and\ \bibinfo {author} {\bibfnamefont {A.~M.}\ \bibnamefont
  {Rey}},\ }\href {\doibase 10.1103/PhysRevA.96.033607} {\bibfield  {journal}
  {\bibinfo  {journal} {Phys. Rev. A}\ }\textbf {\bibinfo {volume} {96}},\
  \bibinfo {pages} {033607} (\bibinfo {year} {2017})}\BibitemShut {NoStop}%
\bibitem [{\citenamefont {Wall}\ \emph {et~al.}(2017)\citenamefont {Wall},
  \citenamefont {Safavi-Naini},\ and\ \citenamefont
  {Rey}}]{PhysRevA.95.013602}%
  \BibitemOpen
  \bibfield  {author} {\bibinfo {author} {\bibfnamefont {M.~L.}\ \bibnamefont
  {Wall}}, \bibinfo {author} {\bibfnamefont {A.}~\bibnamefont {Safavi-Naini}},
  \ and\ \bibinfo {author} {\bibfnamefont {A.~M.}\ \bibnamefont {Rey}},\ }\href
  {\doibase 10.1103/PhysRevA.95.013602} {\bibfield  {journal} {\bibinfo
  {journal} {Phys. Rev. A}\ }\textbf {\bibinfo {volume} {95}},\ \bibinfo
  {pages} {013602} (\bibinfo {year} {2017})}\BibitemShut {NoStop}%
\bibitem [{\citenamefont {G{\"a}rttner}\ \emph {et~al.}(2017)\citenamefont
  {G{\"a}rttner}, \citenamefont {Bohnet}, \citenamefont {Safavi-Naini},
  \citenamefont {Wall}, \citenamefont {Bollinger},\ and\ \citenamefont
  {Rey}}]{Garttner2017}%
  \BibitemOpen
  \bibfield  {author} {\bibinfo {author} {\bibfnamefont {M.}~\bibnamefont
  {G{\"a}rttner}}, \bibinfo {author} {\bibfnamefont {J.~G.}\ \bibnamefont
  {Bohnet}}, \bibinfo {author} {\bibfnamefont {A.}~\bibnamefont
  {Safavi-Naini}}, \bibinfo {author} {\bibfnamefont {M.~L.}\ \bibnamefont
  {Wall}}, \bibinfo {author} {\bibfnamefont {J.~J.}\ \bibnamefont {Bollinger}},
  \ and\ \bibinfo {author} {\bibfnamefont {A.~M.}\ \bibnamefont {Rey}},\ }\href
  {http://dx.doi.org/10.1038/nphys4119} {\bibfield  {journal} {\bibinfo
  {journal} {Nature Physics}\ }\textbf {\bibinfo {volume} {13}},\ \bibinfo
  {pages} {781 EP} (\bibinfo {year} {2017})},\ \bibinfo {note}
  {article}\BibitemShut {NoStop}%
\bibitem [{\citenamefont {Wall}\ \emph {et~al.}(2016)\citenamefont {Wall},
  \citenamefont {Safavi-Naini},\ and\ \citenamefont
  {Rey}}]{PhysRevA.94.053637}%
  \BibitemOpen
  \bibfield  {author} {\bibinfo {author} {\bibfnamefont {M.~L.}\ \bibnamefont
  {Wall}}, \bibinfo {author} {\bibfnamefont {A.}~\bibnamefont {Safavi-Naini}},
  \ and\ \bibinfo {author} {\bibfnamefont {A.~M.}\ \bibnamefont {Rey}},\ }\href
  {\doibase 10.1103/PhysRevA.94.053637} {\bibfield  {journal} {\bibinfo
  {journal} {Phys. Rev. A}\ }\textbf {\bibinfo {volume} {94}},\ \bibinfo
  {pages} {053637} (\bibinfo {year} {2016})}\BibitemShut {NoStop}%
\bibitem [{\citenamefont {Gra{\ss}}\ \emph {et~al.}(2018)\citenamefont
  {Gra{\ss}}, \citenamefont {Celi}, \citenamefont {Pagano},\ and\ \citenamefont
  {Lewenstein}}]{PhysRevA.97.010302}%
  \BibitemOpen
  \bibfield  {author} {\bibinfo {author} {\bibfnamefont {T.}~\bibnamefont
  {Gra{\ss}}}, \bibinfo {author} {\bibfnamefont {A.}~\bibnamefont {Celi}},
  \bibinfo {author} {\bibfnamefont {G.}~\bibnamefont {Pagano}}, \ and\ \bibinfo
  {author} {\bibfnamefont {M.}~\bibnamefont {Lewenstein}},\ }\href {\doibase
  10.1103/PhysRevA.97.010302} {\bibfield  {journal} {\bibinfo  {journal} {Phys.
  Rev. A}\ }\textbf {\bibinfo {volume} {97}},\ \bibinfo {pages} {010302}
  (\bibinfo {year} {2018})}\BibitemShut {NoStop}%
\bibitem [{\citenamefont {Gra{\ss}}\ \emph
  {et~al.}(2016{\natexlab{a}})\citenamefont {Gra{\ss}}, \citenamefont
  {Lewenstein},\ and\ \citenamefont {Bermudez}}]{1367-2630-18-3-033011}%
  \BibitemOpen
  \bibfield  {author} {\bibinfo {author} {\bibfnamefont {T.}~\bibnamefont
  {Gra{\ss}}}, \bibinfo {author} {\bibfnamefont {M.}~\bibnamefont
  {Lewenstein}}, \ and\ \bibinfo {author} {\bibfnamefont {A.}~\bibnamefont
  {Bermudez}},\ }\href {http://stacks.iop.org/1367-2630/18/i=3/a=033011}
  {\bibfield  {journal} {\bibinfo  {journal} {New Journal of Physics}\ }\textbf
  {\bibinfo {volume} {18}},\ \bibinfo {pages} {033011} (\bibinfo {year}
  {2016}{\natexlab{a}})}\BibitemShut {NoStop}%
\bibitem [{\citenamefont {Gra{\ss}}\ \emph {et~al.}(2015)\citenamefont
  {Gra{\ss}}, \citenamefont {Muschik}, \citenamefont {Celi}, \citenamefont
  {Chhajlany},\ and\ \citenamefont {Lewenstein}}]{PhysRevA.91.063612}%
  \BibitemOpen
  \bibfield  {author} {\bibinfo {author} {\bibfnamefont {T.}~\bibnamefont
  {Gra{\ss}}}, \bibinfo {author} {\bibfnamefont {C.}~\bibnamefont {Muschik}},
  \bibinfo {author} {\bibfnamefont {A.}~\bibnamefont {Celi}}, \bibinfo {author}
  {\bibfnamefont {R.~W.}\ \bibnamefont {Chhajlany}}, \ and\ \bibinfo {author}
  {\bibfnamefont {M.}~\bibnamefont {Lewenstein}},\ }\href {\doibase
  10.1103/PhysRevA.91.063612} {\bibfield  {journal} {\bibinfo  {journal} {Phys.
  Rev. A}\ }\textbf {\bibinfo {volume} {91}},\ \bibinfo {pages} {063612}
  (\bibinfo {year} {2015})}\BibitemShut {NoStop}%
\bibitem [{\citenamefont {Nevado}\ \emph {et~al.}(2017)\citenamefont {Nevado},
  \citenamefont {Fern{\'a}ndez-Lorenzo},\ and\ \citenamefont
  {Porras}}]{PhysRevLett.119.210401}%
  \BibitemOpen
  \bibfield  {author} {\bibinfo {author} {\bibfnamefont {P.}~\bibnamefont
  {Nevado}}, \bibinfo {author} {\bibfnamefont {S.}~\bibnamefont
  {Fern{\'a}ndez-Lorenzo}}, \ and\ \bibinfo {author} {\bibfnamefont
  {D.}~\bibnamefont {Porras}},\ }\href {\doibase
  10.1103/PhysRevLett.119.210401} {\bibfield  {journal} {\bibinfo  {journal}
  {Phys. Rev. Lett.}\ }\textbf {\bibinfo {volume} {119}},\ \bibinfo {pages}
  {210401} (\bibinfo {year} {2017})}\BibitemShut {NoStop}%
\bibitem [{\citenamefont {{Nevado}}\ and\ \citenamefont
  {{Porras}}(2015)}]{2015PhRvA..92a3624N}%
  \BibitemOpen
  \bibfield  {author} {\bibinfo {author} {\bibfnamefont {P.}~\bibnamefont
  {{Nevado}}}\ and\ \bibinfo {author} {\bibfnamefont {D.}~\bibnamefont
  {{Porras}}},\ }\href {\doibase 10.1103/PhysRevA.92.013624} {\bibfield
  {journal} {\bibinfo  {journal} {\pra}\ }\textbf {\bibinfo {volume} {92}},\
  \bibinfo {eid} {013624} (\bibinfo {year} {2015})},\ \Eprint
  {http://arxiv.org/abs/1503.04614} {arXiv:1503.04614 [quant-ph]} \BibitemShut
  {NoStop}%
\bibitem [{\citenamefont {Nevado}\ and\ \citenamefont
  {Porras}(2016)}]{PhysRevA.93.013625}%
  \BibitemOpen
  \bibfield  {author} {\bibinfo {author} {\bibfnamefont {P.}~\bibnamefont
  {Nevado}}\ and\ \bibinfo {author} {\bibfnamefont {D.}~\bibnamefont
  {Porras}},\ }\href {\doibase 10.1103/PhysRevA.93.013625} {\bibfield
  {journal} {\bibinfo  {journal} {Phys. Rev. A}\ }\textbf {\bibinfo {volume}
  {93}},\ \bibinfo {pages} {013625} (\bibinfo {year} {2016})}\BibitemShut
  {NoStop}%
\bibitem [{\citenamefont {{Lemmer}}\ \emph {et~al.}(2017)\citenamefont
  {{Lemmer}}, \citenamefont {{Cormick}}, \citenamefont {{Tamascelli}},
  \citenamefont {{Schaetz}}, \citenamefont {{Huelga}},\ and\ \citenamefont
  {{Plenio}}}]{2017arXiv170400629L}%
  \BibitemOpen
  \bibfield  {author} {\bibinfo {author} {\bibfnamefont {A.}~\bibnamefont
  {{Lemmer}}}, \bibinfo {author} {\bibfnamefont {C.}~\bibnamefont {{Cormick}}},
  \bibinfo {author} {\bibfnamefont {D.}~\bibnamefont {{Tamascelli}}}, \bibinfo
  {author} {\bibfnamefont {T.}~\bibnamefont {{Schaetz}}}, \bibinfo {author}
  {\bibfnamefont {S.~F.}\ \bibnamefont {{Huelga}}}, \ and\ \bibinfo {author}
  {\bibfnamefont {M.~B.}\ \bibnamefont {{Plenio}}},\ }\href@noop {} {\bibfield
  {journal} {\bibinfo  {journal} {ArXiv e-prints}\ } (\bibinfo {year}
  {2017})},\ \Eprint {http://arxiv.org/abs/1704.00629} {arXiv:1704.00629
  [quant-ph]} \BibitemShut {NoStop}%
\bibitem [{\citenamefont {Hauke}\ \emph {et~al.}(2015)\citenamefont {Hauke},
  \citenamefont {Bonnes}, \citenamefont {Heyl},\ and\ \citenamefont
  {Lechner}}]{HBHL}%
  \BibitemOpen
  \bibfield  {author} {\bibinfo {author} {\bibfnamefont {P.}~\bibnamefont
  {Hauke}}, \bibinfo {author} {\bibfnamefont {L.~W.}\ \bibnamefont {Bonnes}},
  \bibinfo {author} {\bibfnamefont {M.}~\bibnamefont {Heyl}}, \ and\ \bibinfo
  {author} {\bibfnamefont {W.}~\bibnamefont {Lechner}},\ }\href {\doibase
  10.3389/fphy.2015.00021} {\bibfield  {journal} {\bibinfo  {journal}
  {Frontiers in Physics}\ }\textbf {\bibinfo {volume} {3}},\ \bibinfo {eid}
  {21} (\bibinfo {year} {2015})}\BibitemShut {NoStop}%
\bibitem [{\citenamefont {Gra{\ss}}\ \emph
  {et~al.}(2016{\natexlab{b}})\citenamefont {Gra{\ss}}, \citenamefont
  {Ravent{\'o}s}, \citenamefont {Juli{\'a}-D{\'i}az}, \citenamefont {Gogolin},\
  and\ \citenamefont {Lewenstein}}]{ncomms11524}%
  \BibitemOpen
  \bibfield  {author} {\bibinfo {author} {\bibfnamefont {T.}~\bibnamefont
  {Gra{\ss}}}, \bibinfo {author} {\bibfnamefont {D.}~\bibnamefont
  {Ravent{\'o}s}}, \bibinfo {author} {\bibfnamefont {B.}~\bibnamefont
  {Juli{\'a}-D{\'i}az}}, \bibinfo {author} {\bibfnamefont {C.}~\bibnamefont
  {Gogolin}}, \ and\ \bibinfo {author} {\bibfnamefont {M.}~\bibnamefont
  {Lewenstein}},\ }\href {\doibase 10.1038/ncomms11524} {\bibfield  {journal}
  {\bibinfo  {journal} {Nat. Comms.}\ }\textbf {\bibinfo {volume} {7}}
  (\bibinfo {year} {2016}{\natexlab{b}}),\ 10.1038/ncomms11524}\BibitemShut
  {NoStop}%
\bibitem [{\citenamefont {Barahona}(1982)}]{0305-4470-15-10-028}%
  \BibitemOpen
  \bibfield  {author} {\bibinfo {author} {\bibfnamefont {F.}~\bibnamefont
  {Barahona}},\ }\href {http://stacks.iop.org/0305-4470/15/i=10/a=028}
  {\bibfield  {journal} {\bibinfo  {journal} {Journal of Physics A:
  Mathematical and General}\ }\textbf {\bibinfo {volume} {15}},\ \bibinfo
  {pages} {3241} (\bibinfo {year} {1982})}\BibitemShut {NoStop}%
\bibitem [{\citenamefont {Venuti}\ \emph {et~al.}(2016)\citenamefont {Venuti},
  \citenamefont {Albash}, \citenamefont {Lidar},\ and\ \citenamefont
  {Zanardi}}]{PhysRevA.93.032118}%
  \BibitemOpen
  \bibfield  {author} {\bibinfo {author} {\bibfnamefont {L.~C.}\ \bibnamefont
  {Venuti}}, \bibinfo {author} {\bibfnamefont {T.}~\bibnamefont {Albash}},
  \bibinfo {author} {\bibfnamefont {D.~A.}\ \bibnamefont {Lidar}}, \ and\
  \bibinfo {author} {\bibfnamefont {P.}~\bibnamefont {Zanardi}},\ }\href
  {\doibase 10.1103/PhysRevA.93.032118} {\bibfield  {journal} {\bibinfo
  {journal} {Phys. Rev. A}\ }\textbf {\bibinfo {volume} {93}},\ \bibinfo
  {pages} {032118} (\bibinfo {year} {2016})}\BibitemShut {NoStop}%
\bibitem [{\citenamefont {Nishimura}\ \emph {et~al.}(2016)\citenamefont
  {Nishimura}, \citenamefont {Nishimori}, \citenamefont {Ochoa},\ and\
  \citenamefont {Katzgraber}}]{PhysRevE.94.032105}%
  \BibitemOpen
  \bibfield  {author} {\bibinfo {author} {\bibfnamefont {K.}~\bibnamefont
  {Nishimura}}, \bibinfo {author} {\bibfnamefont {H.}~\bibnamefont
  {Nishimori}}, \bibinfo {author} {\bibfnamefont {A.~J.}\ \bibnamefont
  {Ochoa}}, \ and\ \bibinfo {author} {\bibfnamefont {H.~G.}\ \bibnamefont
  {Katzgraber}},\ }\href {\doibase 10.1103/PhysRevE.94.032105} {\bibfield
  {journal} {\bibinfo  {journal} {Phys. Rev. E}\ }\textbf {\bibinfo {volume}
  {94}},\ \bibinfo {pages} {032105} (\bibinfo {year} {2016})}\BibitemShut
  {NoStop}%
\bibitem [{\citenamefont {Nishimura}\ and\ \citenamefont
  {Nishimori}(2017)}]{PhysRevA.96.042310}%
  \BibitemOpen
  \bibfield  {author} {\bibinfo {author} {\bibfnamefont {K.}~\bibnamefont
  {Nishimura}}\ and\ \bibinfo {author} {\bibfnamefont {H.}~\bibnamefont
  {Nishimori}},\ }\href {\doibase 10.1103/PhysRevA.96.042310} {\bibfield
  {journal} {\bibinfo  {journal} {Phys. Rev. A}\ }\textbf {\bibinfo {volume}
  {96}},\ \bibinfo {pages} {042310} (\bibinfo {year} {2017})}\BibitemShut
  {NoStop}%
\bibitem [{\citenamefont {Chancellor}\ \emph {et~al.}(2016)\citenamefont
  {Chancellor}, \citenamefont {Szoke}, \citenamefont {Vinci}, \citenamefont
  {Aeppli},\ and\ \citenamefont {Warburton}}]{srep22318}%
  \BibitemOpen
  \bibfield  {author} {\bibinfo {author} {\bibfnamefont {N.}~\bibnamefont
  {Chancellor}}, \bibinfo {author} {\bibfnamefont {S.}~\bibnamefont {Szoke}},
  \bibinfo {author} {\bibfnamefont {W.}~\bibnamefont {Vinci}}, \bibinfo
  {author} {\bibfnamefont {G.}~\bibnamefont {Aeppli}}, \ and\ \bibinfo {author}
  {\bibfnamefont {P.~A.}\ \bibnamefont {Warburton}},\ }\href {\doibase
  10.1038/srep22318} {\bibfield  {journal} {\bibinfo  {journal} {Scientific
  Reports}\ }\textbf {\bibinfo {volume} {6}},\ \bibinfo {pages} {22318}
  (\bibinfo {year} {2016})}\BibitemShut {NoStop}%
\bibitem [{\citenamefont {Hairer}\ \emph {et~al.}(1993)\citenamefont {Hairer},
  \citenamefont {N{\o}rsett},\ and\ \citenamefont {Wanner}}]{ODEXcite}%
  \BibitemOpen
  \bibfield  {author} {\bibinfo {author} {\bibfnamefont {E.}~\bibnamefont
  {Hairer}}, \bibinfo {author} {\bibfnamefont {S.~P.}\ \bibnamefont
  {N{\o}rsett}}, \ and\ \bibinfo {author} {\bibfnamefont {G.}~\bibnamefont
  {Wanner}},\ }\href@noop {} {\emph {\bibinfo {title} {{Solving Ordinary
  Differential Equations I. Nonstiff Problems.}}}},\ \bibinfo {edition} {2nd}\
  ed.,\ {Springer Series in Comput. Mathematics}\ (\bibinfo  {publisher}
  {Springer-Verlag},\ \bibinfo {address} {Berlin Heidelberg},\ \bibinfo {year}
  {1993})\BibitemShut {NoStop}%
\bibitem [{\citenamefont {Kim}\ \emph {et~al.}(2009)\citenamefont {Kim},
  \citenamefont {Chang}, \citenamefont {Islam}, \citenamefont {Korenblit},
  \citenamefont {Duan},\ and\ \citenamefont {Monroe}}]{PhysRevLett.103.120502}%
  \BibitemOpen
  \bibfield  {author} {\bibinfo {author} {\bibfnamefont {K.}~\bibnamefont
  {Kim}}, \bibinfo {author} {\bibfnamefont {M.-S.}\ \bibnamefont {Chang}},
  \bibinfo {author} {\bibfnamefont {R.}~\bibnamefont {Islam}}, \bibinfo
  {author} {\bibfnamefont {S.}~\bibnamefont {Korenblit}}, \bibinfo {author}
  {\bibfnamefont {L.-M.}\ \bibnamefont {Duan}}, \ and\ \bibinfo {author}
  {\bibfnamefont {C.}~\bibnamefont {Monroe}},\ }\href {\doibase
  10.1103/PhysRevLett.103.120502} {\bibfield  {journal} {\bibinfo  {journal}
  {Phys. Rev. Lett.}\ }\textbf {\bibinfo {volume} {103}},\ \bibinfo {pages}
  {120502} (\bibinfo {year} {2009})}\BibitemShut {NoStop}%
\bibitem [{\citenamefont {Dickson}\ \emph {et~al.}(2013)\citenamefont
  {Dickson}, \citenamefont {Johnson}, \citenamefont {Amin}, \citenamefont
  {Harris}, \citenamefont {Altomare}, \citenamefont {Berkley}, \citenamefont
  {Bunyk}, \citenamefont {Cai}, \citenamefont {Chapple}, \citenamefont
  {Chavez}, \citenamefont {Cioata}, \citenamefont {Cirip}, \citenamefont
  {deBuen}, \citenamefont {Drew-Brook}, \citenamefont {Enderud}, \citenamefont
  {Gildert}, \citenamefont {Hamze}, \citenamefont {Hilton}, \citenamefont
  {Hoskinson}, \citenamefont {Karimi}, \citenamefont {Ladizinsky},
  \citenamefont {Ladizinsky}, \citenamefont {Lanting}, \citenamefont {Mahon},
  \citenamefont {Neufeld}, \citenamefont {Oh}, \citenamefont {Perminov},
  \citenamefont {Petroff}, \citenamefont {Przybysz}, \citenamefont {Rich},
  \citenamefont {Spear}, \citenamefont {Tcaciuc}, \citenamefont {Thom},
  \citenamefont {Tolkacheva}, \citenamefont {Uchaikin}, \citenamefont {Wang},
  \citenamefont {Wilson}, \citenamefont {Merali},\ and\ \citenamefont
  {Rose}}]{ncomms2920}%
  \BibitemOpen
  \bibfield  {author} {\bibinfo {author} {\bibfnamefont {N.~G.}\ \bibnamefont
  {Dickson}}, \bibinfo {author} {\bibfnamefont {M.~W.}\ \bibnamefont
  {Johnson}}, \bibinfo {author} {\bibfnamefont {M.~H.}\ \bibnamefont {Amin}},
  \bibinfo {author} {\bibfnamefont {R.}~\bibnamefont {Harris}}, \bibinfo
  {author} {\bibfnamefont {F.}~\bibnamefont {Altomare}}, \bibinfo {author}
  {\bibfnamefont {A.~J.}\ \bibnamefont {Berkley}}, \bibinfo {author}
  {\bibfnamefont {P.}~\bibnamefont {Bunyk}}, \bibinfo {author} {\bibfnamefont
  {J.}~\bibnamefont {Cai}}, \bibinfo {author} {\bibfnamefont {E.~M.}\
  \bibnamefont {Chapple}}, \bibinfo {author} {\bibfnamefont {P.}~\bibnamefont
  {Chavez}}, \bibinfo {author} {\bibfnamefont {F.}~\bibnamefont {Cioata}},
  \bibinfo {author} {\bibfnamefont {T.}~\bibnamefont {Cirip}}, \bibinfo
  {author} {\bibfnamefont {P.}~\bibnamefont {deBuen}}, \bibinfo {author}
  {\bibfnamefont {M.}~\bibnamefont {Drew-Brook}}, \bibinfo {author}
  {\bibfnamefont {C.}~\bibnamefont {Enderud}}, \bibinfo {author} {\bibfnamefont
  {S.}~\bibnamefont {Gildert}}, \bibinfo {author} {\bibfnamefont
  {F.}~\bibnamefont {Hamze}}, \bibinfo {author} {\bibfnamefont {J.~P.}\
  \bibnamefont {Hilton}}, \bibinfo {author} {\bibfnamefont {E.}~\bibnamefont
  {Hoskinson}}, \bibinfo {author} {\bibfnamefont {K.}~\bibnamefont {Karimi}},
  \bibinfo {author} {\bibfnamefont {E.}~\bibnamefont {Ladizinsky}}, \bibinfo
  {author} {\bibfnamefont {N.}~\bibnamefont {Ladizinsky}}, \bibinfo {author}
  {\bibfnamefont {T.}~\bibnamefont {Lanting}}, \bibinfo {author} {\bibfnamefont
  {T.}~\bibnamefont {Mahon}}, \bibinfo {author} {\bibfnamefont
  {R.}~\bibnamefont {Neufeld}}, \bibinfo {author} {\bibfnamefont
  {T.}~\bibnamefont {Oh}}, \bibinfo {author} {\bibfnamefont {I.}~\bibnamefont
  {Perminov}}, \bibinfo {author} {\bibfnamefont {C.}~\bibnamefont {Petroff}},
  \bibinfo {author} {\bibfnamefont {A.}~\bibnamefont {Przybysz}}, \bibinfo
  {author} {\bibfnamefont {C.}~\bibnamefont {Rich}}, \bibinfo {author}
  {\bibfnamefont {P.}~\bibnamefont {Spear}}, \bibinfo {author} {\bibfnamefont
  {A.}~\bibnamefont {Tcaciuc}}, \bibinfo {author} {\bibfnamefont {M.~C.}\
  \bibnamefont {Thom}}, \bibinfo {author} {\bibfnamefont {E.}~\bibnamefont
  {Tolkacheva}}, \bibinfo {author} {\bibfnamefont {S.}~\bibnamefont
  {Uchaikin}}, \bibinfo {author} {\bibfnamefont {J.}~\bibnamefont {Wang}},
  \bibinfo {author} {\bibfnamefont {A.~B.}\ \bibnamefont {Wilson}}, \bibinfo
  {author} {\bibfnamefont {Z.}~\bibnamefont {Merali}}, \ and\ \bibinfo {author}
  {\bibfnamefont {G.}~\bibnamefont {Rose}},\ }\href {\doibase
  10.1038/ncomms2920} {\bibfield  {journal} {\bibinfo  {journal} {Nature
  Communications}\ }\textbf {\bibinfo {volume} {4}},\ \bibinfo {pages} {1903}
  (\bibinfo {year} {2013})}\BibitemShut {NoStop}%
\bibitem [{\citenamefont {Roos}\ \emph {et~al.}(2000)\citenamefont {Roos},
  \citenamefont {Leibfried}, \citenamefont {Mundt}, \citenamefont
  {Schmidt-Kaler}, \citenamefont {Eschner},\ and\ \citenamefont
  {Blatt}}]{PhysRevLett.85.5547}%
  \BibitemOpen
  \bibfield  {author} {\bibinfo {author} {\bibfnamefont {C.~F.}\ \bibnamefont
  {Roos}}, \bibinfo {author} {\bibfnamefont {D.}~\bibnamefont {Leibfried}},
  \bibinfo {author} {\bibfnamefont {A.}~\bibnamefont {Mundt}}, \bibinfo
  {author} {\bibfnamefont {F.}~\bibnamefont {Schmidt-Kaler}}, \bibinfo {author}
  {\bibfnamefont {J.}~\bibnamefont {Eschner}}, \ and\ \bibinfo {author}
  {\bibfnamefont {R.}~\bibnamefont {Blatt}},\ }\href {\doibase
  10.1103/PhysRevLett.85.5547} {\bibfield  {journal} {\bibinfo  {journal}
  {Phys. Rev. Lett.}\ }\textbf {\bibinfo {volume} {85}},\ \bibinfo {pages}
  {5547} (\bibinfo {year} {2000})}\BibitemShut {NoStop}%
\bibitem [{\citenamefont {Lin}\ \emph {et~al.}(2013)\citenamefont {Lin},
  \citenamefont {Gaebler}, \citenamefont {Tan}, \citenamefont {Bowler},
  \citenamefont {Jost}, \citenamefont {Leibfried},\ and\ \citenamefont
  {Wineland}}]{PhysRevLett.110.153002}%
  \BibitemOpen
  \bibfield  {author} {\bibinfo {author} {\bibfnamefont {Y.}~\bibnamefont
  {Lin}}, \bibinfo {author} {\bibfnamefont {J.~P.}\ \bibnamefont {Gaebler}},
  \bibinfo {author} {\bibfnamefont {T.~R.}\ \bibnamefont {Tan}}, \bibinfo
  {author} {\bibfnamefont {R.}~\bibnamefont {Bowler}}, \bibinfo {author}
  {\bibfnamefont {J.~D.}\ \bibnamefont {Jost}}, \bibinfo {author}
  {\bibfnamefont {D.}~\bibnamefont {Leibfried}}, \ and\ \bibinfo {author}
  {\bibfnamefont {D.~J.}\ \bibnamefont {Wineland}},\ }\href {\doibase
  10.1103/PhysRevLett.110.153002} {\bibfield  {journal} {\bibinfo  {journal}
  {Phys. Rev. Lett.}\ }\textbf {\bibinfo {volume} {110}},\ \bibinfo {pages}
  {153002} (\bibinfo {year} {2013})}\BibitemShut {NoStop}%
\bibitem [{\citenamefont {Lechner}\ \emph {et~al.}(2016)\citenamefont
  {Lechner}, \citenamefont {Maier}, \citenamefont {Hempel}, \citenamefont
  {Jurcevic}, \citenamefont {Lanyon}, \citenamefont {Monz}, \citenamefont
  {Brownnutt}, \citenamefont {Blatt},\ and\ \citenamefont
  {Roos}}]{PhysRevA.93.053401}%
  \BibitemOpen
  \bibfield  {author} {\bibinfo {author} {\bibfnamefont {R.}~\bibnamefont
  {Lechner}}, \bibinfo {author} {\bibfnamefont {C.}~\bibnamefont {Maier}},
  \bibinfo {author} {\bibfnamefont {C.}~\bibnamefont {Hempel}}, \bibinfo
  {author} {\bibfnamefont {P.}~\bibnamefont {Jurcevic}}, \bibinfo {author}
  {\bibfnamefont {B.~P.}\ \bibnamefont {Lanyon}}, \bibinfo {author}
  {\bibfnamefont {T.}~\bibnamefont {Monz}}, \bibinfo {author} {\bibfnamefont
  {M.}~\bibnamefont {Brownnutt}}, \bibinfo {author} {\bibfnamefont
  {R.}~\bibnamefont {Blatt}}, \ and\ \bibinfo {author} {\bibfnamefont {C.~F.}\
  \bibnamefont {Roos}},\ }\href {\doibase 10.1103/PhysRevA.93.053401}
  {\bibfield  {journal} {\bibinfo  {journal} {Phys. Rev. A}\ }\textbf {\bibinfo
  {volume} {93}},\ \bibinfo {pages} {053401} (\bibinfo {year}
  {2016})}\BibitemShut {NoStop}%
\end{thebibliography}%

\appendix:
\begin{figure}[t]
\includegraphics[width=0.5\textwidth]{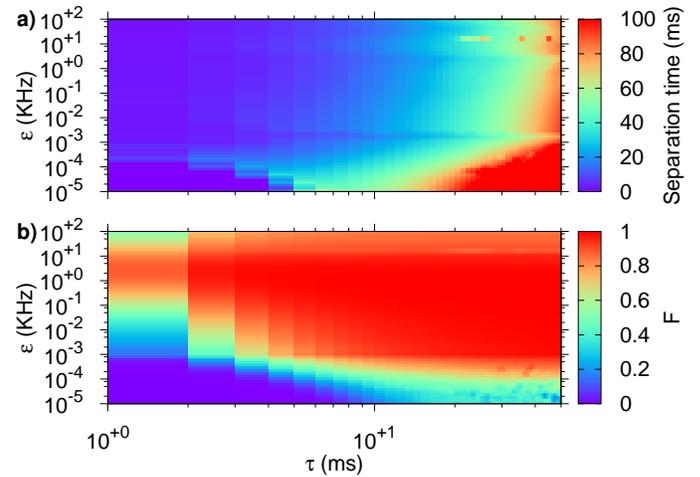}
\caption{\label{fig:optim_bias} Separation time (a)) and Fidelity (b)) 
of systems with $6$ spins for several values of $\varepsilon$ and 
$\tau$, with initial populations of phononic modes set to $0$. 
Waiting time is fixed to $20 \tau$ and $\omega_L = \omega_{\rm N}-1500$ kHz.}
\end{figure}

\section{Robustness of the semiclassical calculation}

 \label{app:Integrationtests}

One can test the quality of the integration method by checking whether the constants 
of motion are conserved.  In time-independent systems, the total energy of the 
system is ususally the conserved quantity which can be computs most easily. 
Unfortunately, since the Hamiltonian in Eq.~\ref{eq:Hamiltonian} does not commute 
with itself at different times due to the transverse magnetic field in the annealing 
term, the total energy of the system is not conserved. Furthermore, it is challenging 
to find an analytical expression for an alternative conserved quantity 
given the infinitely non-commutativity of the Pauli matrices algebra.
We have tested the integration method checking its time reversibility.
The latter test has given relative differences below $10^-7$, which is within the range of the computation precision.

\section{Optimal bias for the exponential annealing function}
\label{sec:optim}

Although the only function of the bias potential $\epsilon$ is to break $\mathcal{Z}_2$ symmetry in the target Hamiltonian, it turns out that the value and position of the bias have a non-neglible effect on the outcome of the annealing process. Here, we investigate which values of the bias $\epsilon$  and the annealing parameter $\tau$ 
minimize the separation time. As seen in 
Fig.~\ref{fig:optim_bias} a), the separation time is minimized for smaller values 
of $\tau$, at any value of the bias larger than Hz. However, small $\tau$ are known to affect negatively the
fidelity. As we see from Fig.~\ref{fig:optim_bias} b), there is, even for decay times as short as a few ms, a range of bias potentials (roughly between 1 kHz and 10 kHz), where the fidelity gets large. Thus, this range defines the optimal choice for $\epsilon$, which we have also used in our calculations.

We also note that the magnitude of the bias provides a bound for the maximum absolute value of $\langle \sigma_{x} \rangle$, that is, the 
spin expectation on the biased site at the end of the annealing depends on the strength of the bias potential. For a weak bias, this spin will deviate only weakly from zero, limiting the overall fidelity. 
\end{document}